\documentclass[11pt]{article}
\usepackage{amsmath, amssymb, graphics}

\usepackage[nosort]{cite}
\newcommand{\mathsym}[1]{{}}
\usepackage{verbatim}
\usepackage{amsfonts,euscript,amssymb,stmaryrd,braket}
\usepackage{graphics,tikz}
\usetikzlibrary{arrows,decorations.markings,patterns}
\usepackage{caption}
\usepackage{subcaption}
\usepackage{slashed}
\definecolor{hyperref}{RGB}{026,028,185}
\usepackage[bookmarks=true,colorlinks=true,linkcolor=hyperref,citecolor=hyperref,urlcolor=hyperref,bookmarksnumbered]{hyperref}
 \usepackage{booktabs}
 \usepackage{multirow}
\usepackage{tabularx}
\usepackage{longtable}
\usepackage{bbold,bbding}
\usepackage{amsthm} 
\usepackage{esint}
\newcommand{\bal}{\begin{equation}\begin{aligned}}
\newcommand{\eal}{\end{aligned} \end{equation}}
\def\id{\protect{{1 \kern-.28em {\rm l}}}}

\makeatletter
\renewcommand\section{\@startsection {section}{1}{\z@}%
                                   {-3.5ex \@plus -1ex \@minus -.2ex}%
                                   {2.3ex \@plus.2ex}%
                                   {\normalfont\large\bfseries}}
\renewcommand\subsection{\@startsection{subsection}{2}{\z@}%
                                   {-3.25ex\@plus -1ex \@minus -.2ex}%
                                   {1.5ex \@plus .2ex}%
                                   {\normalfont\normalsize\bfseries}}

\makeatother

\numberwithin{equation}{section}

\usepackage[utf8]{inputenc}
\usepackage{epsfig}
\usepackage{graphicx}
\usepackage[multiple]{footmisc}
\usepackage{amssymb,amsmath}
\usepackage{fancybox,framed,tikz}
\usetikzlibrary{decorations.pathmorphing,patterns}
\usepackage{dsfont}
\usepackage{mathtools}
\usepackage{braket}
\usepackage{slashed}
\usepackage{rotating}
\usepackage{bbold,amsfonts}
\usepackage{multirow}
\usepackage{verbatim}
\usepackage{upgreek}

\newcommand{\beq}{\begin{equation}}
\newcommand{\eeq}{\end{equation}}

\renewcommand{\a}{\alpha}
\renewcommand{\b}{\beta}

\renewcommand{\d}{\delta}
\newcommand{\pa}{\partial}
\newcommand{\g}{\gamma}

\newcommand{\D}{\Delta}
\newcommand{\e}{\epsilon}

\renewcommand{\k}{\kappa}

\renewcommand{\L}{\Lambda}
\newcommand{\m}{\mu}

\newcommand{\n}{\nu}

\newcommand{\p}{\pi}

\newcommand{\s}{\sigma}

\renewcommand{\O}{\Omega}

\newcommand{\Tr}{\textup{Tr}}
\newcommand{\Str}{\textup{Str}}
\newcommand\To{\rule{0pt}{2.6ex}}       
\newcommand\sbullet[1][.5]{\mathbin{\vcenter{\hbox{\scalebox{#1}{$\bullet$}}}}}

\usepackage{color}

\definecolor{mypink1}{rgb}{0.958, 0.188, 0.478}

\newcommand{\ads}{\textup{\textrm{AdS}}}

\newcommand{\sphere}{\textup{\textrm{S}}}

\newcommand{\cp}{\C \textup{\textrm{P}}}
\newcommand{\C}{\mathbb{C}}
\DeclareMathOperator{\vol}{vol}
\newcommand{\la}{\label}
\newcommand{\be}{\begin{eqnarray}}
\newcommand{\ee}{\end{eqnarray}}
\def\ov{\over}
\def \del{\partial}
\DeclarePairedDelimiter{\abs}{\lvert}{\rvert}
  \def \tet {\textstyle}
 \def\wb{\bar w}
 \def\r{\rho}
 \def\n{\nu}
 \def\Xb{\bar X}
 \def\k{\kappa}

\hypersetup{}

\tikzset{Witten diagram/.style={execute at begin picture={%
\draw[blue ,fill=blue!05] circle[radius=\pgfkeysvalueof{/tikz/Witten/radius}];
\path node (X){\phantom{X}};
},baseline={(X.base)}},vertex/.style={circle,fill,inner sep=1.414pt,node
contents={}},
Witten/.cd,radius/.initial=1.414cm}
\newenvironment{wittendiagram}[1][]{\begin{tikzpicture}[Witten diagram,#1]}{\end{tikzpicture}}

\DeclarePairedDelimiter{\cor}{\langle}{\rangle}

\begin{document}
\renewcommand{\thefootnote}{\arabic{footnote}}

\overfullrule=0pt
\parskip=2pt
\parindent=12pt
\headheight=0in \headsep=0in \topmargin=0in \oddsidemargin=0in

\vspace{ -3cm} \thispagestyle{empty} \vspace{-1cm}
\begin{flushright} 
\footnotesize
{HU-EP-20/07}
 
\end{flushright}%

\begin{center}
\vspace{1.2cm}
{\Large\bf \mathversion{bold}
{Analytic bootstrap and Witten diagrams \\for the ABJM Wilson line as defect CFT$_1$}\\
}
 
\author{ABC\thanks{XYZ} \and DEF\thanks{UVW} \and GHI\thanks{XYZ}}
 \vspace{0.8cm} {
 L.~Bianchi$^{a,c}$\footnote{{\tt lorenzo.bianchi@qmul.ac.uk}}, G.~Bliard$^{a,b}$\footnote{{\tt gabriel.bliard@physik.hu-berlin.de}},~V.~Forini$^{a,b}$\footnote{{\tt valentina.forini@city.ac.uk}}, L.~Griguolo$^{d,}$\footnote{{\tt luca.griguolo@fis.unipr.it}}, D.~Seminara$^{e,}$\footnote{{\tt seminara@fi.infn.it}}}
 \vskip  0.5cm

\small
{\em
$^{a}$ Department of Mathematics, City, University of London,\\
Northampton Square, EC1V 0HB London, United Kingdom
\vskip 0.02cm
  $^{b}$  
Institut f\"ur Physik, Humboldt-Universit\"at zu Berlin and IRIS Adlershof, \\Zum Gro\ss en Windkanal 6, 12489 Berlin, Germany    
\vskip 0.02cm
$^{c}$  Centre for Research in String Theory, 
Queen Mary University of London\\ Mile End Road, London E1 4NS, United Kingdom
\vskip 0.02cm
$^{d}$ Dipartimento SMFI, Universit\`a di Parma and INFN Gruppo Collegato di Parma, \\Viale G.P. Usberti 7/A, 43100 Parma, Italy
\vskip 0.02cm
$^{e}$ Dipartimento di Fisica, Universit\`a di Firenze and INFN Sezione di Firenze, \\Via G. Sansone 1, 50019 Sesto Fiorentino, Italy
}
\normalsize

\end{center}

\vspace{0.3cm}
\begin{abstract} 


We study local operator insertions on 1/2-BPS line defects in  ABJM theory. Specifically, we consider a class of four-point correlators in the CFT$_1$ with SU$(1, 1|3)$ superconformal symmetry defined on the 1/2-BPS Wilson line. The relevant insertions belong to the short supermultiplet containing the displacement operator and correspond to fluctuations of the dual fundamental string in $AdS_4 \times \cp^3$ ending on the line at the boundary. We use superspace techniques to represent the displacement supermultiplet and we show that superconformal symmetry determines the four-point correlators of its components in terms of a single function of the one-dimensional cross-ratio. Such function is highly constrained by crossing and internal consistency, allowing us to use an analytical bootstrap approach to find the first subleading correction at strong coupling. Finally, we use AdS/CFT to compute the same four-point functions through tree-level $AdS_2$ Witten diagrams, producing a result that is perfectly consistent with the bootstrap solution.

\end{abstract}

\newpage

\tableofcontents
   
\section{Introduction and Discussion}

Wilson loops are fundamental non-local observables of any gauge theory, and admit a representation in terms of the usual lagrangian fields employed in the weak coupling description. At strong coupling their properties are naturally encoded into the degrees of freedom of a semiclassical open string, when a gauge/gravity description is available \cite{Maldacena:1998im,Rey:1998ik,Drukker:1999zq}. Wilson lines are also a prototypical example of defect in QFT, and could support a defect field theory characterising their dynamical behaviour. In the supersymmetric case, BPS Wilson lines provide one-dimensional supersymmetric defect field theories, explicitly defined through the correlation functions of local operator insertions on the contour~\cite{Drukker:2006xg}. From this perspective, 1/2 BPS Wilson lines in the four-dimensional ${\cal N}=4$ supersymmetric Yang-Mills theory (SYM) have been actively studied in the last few years~\cite{Giombi, Giombi:2018qox,Giombi:2018hsx}. In this case, the 
associated defect field theory is conformal (DCFT). Correlation functions can be generated through a "wavy line" procedure and studied at weak coupling~\cite{Cooke:2017qgm} using general results for Wilson loops~\cite{Bassetto:2008yf}. 
Further information has been gained by considering four-point correlators of certain protected operator insertions~\cite{Giombi} whose two-point functions control the ${\cal N}=4$ SYM Bremsstrahlung function~\cite{Correa:2012at}.  
At strong coupling, these correlators are evaluated studying the relevant AdS/CFT dual string sigma-model. The latter corresponds to an effective field theory in $AdS_2$, and correlations functions can be computed by means of standard Witten diagrams~\cite{Giombi}. 
 The conformal bootstrap has been also applied to the computation of the same four-point functions~\cite{Carlo}, recovering and extending the analytical results at strong coupling and studying numerically the finite-coupling regime. Notably, the same approach led to analogous results in a less supersymmetric scenario~\cite{Gimenez-Grau:2019hez}.   More generally, line defects provide a useful and physically interesting laboratory for the application of the analytical techniques developed in the context of one-dimensional CFTs~\cite{Mazac:2016qev,Maldacena:2016hyu,Simmons-Duffin:2017nub,Mazac:2018mdx,Mazac:2018ycv,Ferrero:2019luz}.
Finite-coupling results on defect conformal data were also obtained using integrability via the quantum spectral curve technology~\cite{Grabner:2020nis}. 

Here we study the DCFT associated to the 1/2 BPS Wilson line in ${\cal N}=6$ Super Chern-Simons theory with matter (ABJM)~\cite{ABJM}. 
The structure of Wilson loops in ABJM theory is richer than in ${\cal N}=4$ SYM~\cite{Drukker:2019bev}, admitting different realizations through the fundamental lagrangian fields~\cite{Drukker:2008zx,Chen:2008bp,Drukker:2009hy}, sometimes leading to the same quantum expectations values through a cohomological equivalence~\cite{Drukker:2009hy}. The relevant defect field theories should be able to fully distinguish them, possibly describing different Wilson lines in terms of marginal deformations~\cite{Correa:2019rdk}. Further motivations are the potential existence of topological sectors, that could be associated to new supersymmetric localization procedures, and the relations with integrability, that might lead to an alternative derivation of the elusive $h(\lambda)$ function of ABJM~\cite{Gromov:2014eha} (see also~\cite{Bianchi:2014ada}). 

\bigskip

In the following we shall be interested in the calculation of defect correlation functions, i.e. correlators of local operators inserted along the 1/2 BPS Wilson line. Given some local operators $\mathcal{O}_i(t_i)$, one can define the gauge invariant Wilson line with insertions as
\begin{equation}\label{WOOO}
\mathcal{W}[\mathcal{O}_1(t_1)\mathcal{O}_2(t_2)\,...\,\mathcal{O}_n(t_n)]\equiv 
\Tr \mathcal{P} \left[\mathcal{W}_{t_i,t_1}\mathcal{O}_1(t_1)\mathcal{W}_{t_1,t_2}\mathcal{O}_2(t_2)\,...\,\mathcal{O}_n(t_n)\mathcal{W}_{t_n,t_f}\right],
\end{equation}
where $t$ parameterizes an infinite straight line and $\mathcal{W}_{t_a,t_b}$ is the path-ordered exponential of a suitable connection, that starts at position $x(t_a)$ and ends at position $x(t_b)$.  We choose $t_i=-\infty$ and $t_f=\infty$. The  local operators $\mathcal{O}_i(t_i)$ are inserted between (untraced) Wilson lines, and therefore are not invariant, but have to transform in the adjoint representation of the gauge group. The (one-dimensional) defect  correlators are then defined as
\begin{equation}\label{npoint}
\langle\mathcal{O}_1(t_1)\mathcal{O}_2(t_2)\,...\,\mathcal{O}_n(t_n)\rangle_{\mathcal{W}}\equiv
\frac{\langle\mathcal{W}[\mathcal{O}_1(t_1)\mathcal{O}_2(t_2)\,...\,\mathcal{O}_n(t_n)]\rangle}{\langle\mathcal{W}\rangle}\,\,.
\end{equation}
This definition of correlators is actually more general than the specific realization in terms of Wilson lines, and it extends to any one-dimensional defect conformal field theory (DCFT).   
Only part of the original  ABJM symmetry OSp$(6|4)$ is preserved: the unbroken supergroup is SU$(1,1|3)$, whose bosonic subgroup is SU$(1,1) \times$  SU$(3)_R \times$ U$(1)_{J_0}$. Defect operators are classified by a set of four quantum numbers $[\Delta, j_0, j_1, j_2]$ associated to the four Cartan generators of this bosonic subalgebra. The structure of short and long multiplets representing this subalgebra has been studied thoroughly  in~\cite{defectABJM} and will be reviewed in the main body of the paper. 

Our study will concentrate mainly on correlators associated to the components of a short multiplet, the displacement multiplet, which plays a fundamental role in any supersymmetric DCFT. It contains the displacement operator $\mathbb{D}(t)$, that is supported on every conformal defect~\cite{Billo:2016cpy} and describes  infinitesimal deformations of the defect profile, as well as other operators associated to the broken R-symmetries and supercharges. 
We will consider three different, complementary realizations of this protected multiplet, that are useful to probe its properties and to calculate its correlation functions, depending on the coupling regime and the computational method. A first,  more general, realization of the displacement multiplet is obtained in terms of a superfield $\Phi$ associated to a short multiplet of SU$(1,1|3)$. A representation theory analysis shows that this superfield is neutral under SU$(3)$ and annihilated by half of the supercharges (and thus chiral or antichiral). In four dimensions, the analogous superfield representation for the displacement multiplet has been derived in~\cite{Carlo}, resulting in a superfield charged under the residual $R$-symmetry group. A natural strong-coupling realization  is  provided by the AdS/CFT correspondence: in ABJM theory a 1/2 BPS Wilson line is dual to the fundamental open string living in $AdS_4\times \cp^3$, with the appropriate boundary conditions on the straight contour. The minimal surface spanned by the string encodes its vacuum expectation value at leading order in the strong-coupling expansion. Fluctuations around the minimal surface solution nicely organize  in a AdS$_2$ multiplet of transverse modes~\cite{Forini:2012bb, Aguilera-Damia}, whose components precisely match the quantum numbers of the displacement multiplet. A summary of this correspondence is provided in Table \ref{tabledisp}. A third realization is obtained by inserting field operators, constructed explicitly from the elementary fields appearing in the ABJM lagrangian. In our case the identification is trickier than in the ${\cal N}=4$ SYM case: the 1/2 BPS Wilson line in ABJM is constructed as the holonomy of a superconnection living on a $U(N|N)$ superalgebra~\cite{Drukker:2009hy, Cardinali:2012ru}, and our multiplet should be represented by the insertion of appropriate supermatrices. Perturbative weak-coupling results for the correlation functions could be obtained in this framework by ordinary Feynman diagrams. We construct explicitly the supermultiplet in terms of supermatrices, taking into account the fact that the action of the relevant supercharges is deformed by the presence of the Wilson line itself~\footnote{A basic difference between 1/2 BPS Wilson lines in ${\cal N}=4$ SYM and ABJM is that in the first case the relevant connection is invariant under supersymmetry, while in the second one it undergoes a supergauge $U(N|N)$ transformation \cite{Drukker:2009hy, Cardinali:2012ru}. This fact will be important in deriving the correct field-theoretical representation of the displacement supermultiplet.}. 
We will use the first two realizations of the displacement supermultiplet in the computation of the four-point correlation functions at strong coupling, verifying their relative consistency. We leave an analogous weak-coupling calculation to future investigations.

\begin{table}
\begin{center}
\begin{tabular}{l c c c}
 \sc Grading &\sc Operator & $\D$ & $m^2$\\
 \hline
 Fermion & $\mathbb{F}(t)$ & $\frac12$ & 0\\
 Boson   & $\mathbb{O}^a(t)$ & $1$ & 0\\
 Fermion & $\mathbb{\Lambda}_a(t)$ & $\frac32$ & 1\\
 Boson   & $\mathbb{D}(t)$ & $2$ & 2\\
 \hline
\end{tabular}
\caption{Components of the displacement supermultiplet with their scaling dimensions and the masses of the dual $AdS_2$ string excitations. The mass is obtained through the AdS/CFT dictionary $m^2=\D(\D-1)$ for the bosons and $m^2=(\Delta-\frac12)^2$ for the fermions.}
\label{tabledisp}
\end{center}
\end{table}

\subsection*{Results}

A first important result of our analysis concerns the general structure of the four-point functions of the displacement supermultiplet. The superfield formalism, together the underlying superconformal symmetry, enables us to determine all the non-vanishing correlators for a given ordering of the external superfields in terms of a single function $f(z)$
of the relevant conformal cross-ratio $z=\frac{t_{12} t_{34}}{t_{13} t_{24}}$. 
This function appears directly in the correlator of the superconformal primary $\mathbb{F}(t)$ and its conjugate $\bar{\mathbb{F}}(t)$ \begin{align}\label{corrFintro}
\braket{\mathbb{F}(t_1)\bar{\mathbb{F}}(t_2)\mathbb{F}(t_3)\bar{\mathbb{F}}(t_4)}&=\frac{C_{\Phi}^2}{t_{12} t_{34}} f(z)\,.
\end{align}
Above, $C_{\Phi}$ is the normalization of the superfield two-point function and has a physical interpretation in terms of the Bremsstrahlung function which will be discussed below, in Section~\ref{sec:displmultiplet}.
In one dimensional CFTs, correlators come with a specific ordering and one is allowed to take OPEs only for neighbouring operators. 
Therefore, crossing symmetry implies that the exchange $1\leftrightarrow 3$ is a symmetry of the correlator~\eqref{corrFintro}, whereas $1\leftrightarrow 2$ is not. This means that there could be independent functions of the cross-ratio associated to different operator orderings, see Section \ref{sec:superspace} for a thorough discussion of this issue. 

Having encoded all the information into this function, we use the analytic bootstrap to compute $f(z)$  in a first-order perturbation around the generalized free-field theory result obtained by Wick contractions. As a first step we carefully examine the OPE structure for the superfield $\Phi$. In our case,  there are  two qualitatively different OPE channels to consider, depending on whether we take the chiral-antichiral OPE $\Phi\times \bar\Phi$  or the chiral-chiral OPE $\Phi \times\Phi$. We have derived the selection rules for the superconformal representations appearing in these two channels, as well as the corresponding superconformal blocks~\footnote{Our analysis holds for a chiral short multiplet whose superprimary has generic $U(1)$ charge.}. In the chiral-antichiral OPE only long multiplets appear and the associated superblocks are explicitly obtained by diagonalizing the superconformal Casimir operator. The chiral-chiral OPE is richer and we observe the presence of three short multiplets in addition to the long ones: importantly the infinitely many long operators appearing here have (unprotected) dimensions strictly higher than the short (protected) ones. This bound is crucial in the solution of the bootstrap equations. After having settled the relevant (super)block expansions, we impose symmetries and consistency with the OPE's, obtaining an infinite family of solutions. We then use a physical criterium to classify these solutions and we select the ``minimal'' one leading to a function $f(z)$
\begin{align}\label{f-intro}
f(z)&=1-z+{\epsilon} \Big[z-1+z (3-z) \log (-z)-\frac{(1-z)^3 }{z}\log (1-z)\Big]+\mathcal{O}(\e^2)\,, & \epsilon&=\frac{1}{4\pi T}\,.
\end{align}
The parameter $\epsilon$ controls the expansion around generalized free-field theory, and it will be interpreted as the inverse string tension  $T$ appearing in the effective AdS$_2$ sigma-model~\footnote{The precise relation between the $\epsilon$ parameter and the string tension $T$ can only be established after the comparison with the explicit Witten diagram computation.}. We then extract the anomalous dimensions and the OPE coefficients of the composite operators appearing in the intermediate channels. At leading order (generalized free-field theory) the operators exchanged in the OPE channels are ``two-particle'' operators of the schematic form $\mathbb{F}\partial^{n}_t\bar{\mathbb{F}}$ in the chiral-anchiral channel, and $\mathbb{F}\partial^{n}_t{\mathbb{F}}$ (with odd $n$) in the chiral-chiral channel~\footnote{At strong coupling these operators should represent worldsheet bound states, made of two of the corresponding fluctuations, as discussed in~\cite{Giombi}.}. 
 We consider the following  perturbation  over their classical dimension
\begin{align}
&\text{chiral-antichiral channel}:& \Delta_n&=1+n+\e\gamma^{(1)}_n+\mathcal{O}(\e^2)\\\label{an-dim-intro}
&\text{chiral-chiral channel}:& \Updelta_n&=1+n+\e \upgamma^{(1)}_n + \mathcal{O}(\e^2)& &n \text{ odd}
\end{align}
and comparing~\eqref{f-intro} with the associated block expansion,  we find the following expression for the anomalous dimensions
\be\label{andim-intro}
\gamma^{(1)}_n=-n^2-4n-3~,\qquad\qquad  \upgamma^{(1)}_n=-n^2-n+2\,,   \qquad n \text{ odd}\,\,,
\ee
and an analogous result for the OPE coefficients. One would be tempted to interpret these formulas as the leading corrections to the classical dimension of the two-particle operators defined above (or their supersymmetric generalization). However, as we discuss in Section~\ref{sec:bootstrap}, two-particle operators mix, in general, with multi-particle operators, and our result should correspond to linear combinations of the actual anomalous dimensions weighted by OPE coefficients~\footnote{There is of course the possibility that, at the first non-trivial order, the degeneracy is not lifted. In this case, the result~\eqref{andim-intro} would provide the eigenvalues of the dilatation operator.}. We stress anyway that, since we bootstrap directly the correlator~\eqref{corrFintro}, the result~\eqref{f-intro} is not affected by mixing. 


As mentioned above, the worldsheet fluctuations around the  (AdS$_2$) minimal surface corresponding to the 1/2 BPS Wilson line are in direct correspondence with the components of the displacement supermultiplet. Through AdS/CFT,  correlators of these  AdS$_2$ fields evaluated at the boundary correspond to correlation functions of the dual defect operators~\cite{Giombi}. For large string tension $T$, their boundary-to-boundary propagator is free, leading to a generalized free-field theory result for their four-point function.  The $1/T$ expansion for the Nambu-Goto string action  involves non-trivial bulk interactions, and the associated boundary correlators are evaluated via AdS$_2$ Witten diagrams~\footnote{We remark that, compared to the usual $1/N$ expansion in Witten diagrams for higher-dimensional AdS/CFT, we are expanding the large-$N$ string sigma-model in inverse powers of the string tension. See related discussion in~\cite{Giombi}.}.   
We derived the effective quartic Lagrangian governing the interactions of the AdS$_2$ fields and obtained the associated Feynman rules. 
The computation of all bosonic correlators confirms the functional form of $f(z)$, in perfect harmony with the bootstrap result once we identify the two expansion parameters as in~\eqref{f-intro}.  

It is interesting to point out similarities and differences between the case of interest in this paper and its four-dimensional counterpart~\cite{Carlo,Giombi}. The structure of the displacement multiplet  for example is different: the superprimary is a fermion, a feature that from the one-dimensional point of view simply amounts to give a Gra\ss mann character to the related field. The three-dimensional representation of the supermultiplet, in terms of Lagrangian fields inserted into the Wilson line, is instead far from being trivial, and implies a sophisticated construction in terms of supermatrices. 
At strong coupling, the fermionic nature of the superprimary implies that the bosonic AdS$_2$ excitations are dual to super-descendants in the displacement supermultiplet. Therefore, their correlators do not provide a direct result for the function $f(z)$, which can be however obtained by comparing the superfield expansion of the correlator with the Witten diagram computation.  This results in a system of differential equations, whose unique solution -- $f(z)$ in~\eqref{f-intro} --  provides a non-trivial consistency check of our procedure.   

Another difference concerns the R-symmetry structure, since the chiral superfield $\Phi$ is neutral under SU$(3)$ and its four-point function does not require any R-symmetry cross-ratio. This prevents the possibility to construct topological operators, whose correlation functions on the line do not depend on the positions of the insertion. This is in sharp contrast with the $\mathcal{N}=4$ case, where topological operators in the displacement multiplet have been found \cite{Giombi,Carlo} and their correlation functions have been computed exactly by localization~\cite{Giombi:2009ds, Giombi:2018qox,Giombi:2018hsx}.  
In our setting topological operators seem instead to appear inside another multiplet~\cite{toappear}, making difficult to connect our computations to some all order result. 

\subsection*{Outlook}

A natural development of this work would be to calculate the four-point functions of the displacement supermultiplet beyond tree-level at strong coupling, using loop corrections to Witten diagrams in AdS$_2$. The relevant AdS$_2$ sigma-model should be UV finite~\cite{Giombi}, but regularization subtleties are anyway expected  in AdS$_2$ models with derivative interactions (for example, see discussion in~\cite{Menotti:2005fk,Menotti:2006tc,Beccaria:2019stp,Beccaria:2019mev,Beccaria:2019dju,Beccaria:2020qtk}). A parallel attempt would be to compute the anomalous dimensions of exchanged operators beyond the first non-trivial order using the bootstrap approach: the potential mixing problem discussed above is expected to arise at this level, and its resolution would require the analysis of different correlators~\footnote{We thank Carlo Meneghelli for discussing with us this possibility.}. Another very interesting direction could be to apply integrability in this context, as done recently in the ${\cal N}=4$ case~\cite{Grabner:2020nis}.  Data at finite coupling for the correlators of interest here may also be obtained with lattice field theory methods applied to the string worldsheet, discretizing the Lagrangian of~\cite{Uvarov:2009nk} expanded around the  minimal surface corresponding to the 1/2 BPS line, and using Monte Carlo techniques on the lines of~\cite{Bianchi:2016cyv, Forini:2016sot, Forini:2017mpu,Forini:2017ene, Bianchi:2019ygz} for the correlators of the worldsheet excitations.  
Weak coupling computations represent also a viable extension of the present work. A traditional perturbative field-theoretical calculation of the correlators, using the supermatrix representation of the displacement multiplet~\footnote{See Section \ref{dispmult}.}, 
should determine the function $f(z)$ at small coupling. It can be done by generalising the procedure developed in~\cite{defectABJM} for the Bremsstrahlung  function and exploiting the Feynman diagrams experience gained in~\cite{Bianchi:2018bke}.  
Topological sectors, hopefully amenable to localization, could  appear considering other supermultiplets~\cite{toappear}: if this is the case, the study of more general correlators might be interesting. It would be also interesting to extend these investigations to non-supersymmetric lines in ABJM, as done in~\cite{Beccaria:2019dws} for ${\cal N}=4$ SYM, or to higher-dimensional defects~\cite{Drukker:2020swu}.  
 
 This paper proceeds as follows. In Section~\ref{sec:superconformal} we discuss the displacement supermultiplet in ABJM, relating its components to the symmetries broken by the line defect. We present its properties from the representation theory point of view, and derive the field theoretical realization of its components as supermatrix-valued insertions in the 1/2 BPS Wilson line. Section~\ref{sec:superspace} is devoted to the chiral superfield approach to the study of the four-point correlation functions, and to the discussion of the different OPE's and selection rules relevant for the bootstrap approach.  The actual derivation of the functions $f(z)$ is contained in Section~\ref{sec:bootstrap}, where the full bootstrap machinery is applied to the four-point correlators and the extraction of the conformal data is discussed. In Section~\ref{sec:sigmamodel} we turn our attention to the computation of the correlators performed via Witten diagrams.  
 A number of technical appendices complete our manuscript.

\section{1/2 BPS Wilson line and the displacement supermultiplet in ABJM}
\label{sec:superconformal}
This section is devoted to the definition and construction of the displacement supermultiplet 
for a line defect given by the $1/2$ BPS Wilson line in ABJM. For completeness, we start by recapitulating
 some basic facts  about this theory. The gauge sector
consist of two gauge fields $A_\mu$ and  $\hat A_\mu$ belonging respectively to 
the adjoint of $U(N)$ and $\hat U(N)$. The matter sector instead contains the complex scalar fields  
$C_I$ and $\bar C^I$ as well as  the fermions $\psi_I$ and $\bar \psi^I$. The fields $(C_I, \bar \psi^I)$ transforms
in the bifundamental  $(N,\bar N)$ while the couple $(\bar C^I,  \psi_I)$ lives in the $(\bar N,N)$.
The additional capital index $I = 1,2,3,4$ label the (anti)fundamental representation of the R-symmetry group SU$(4)$.  The kinetic term for the gauge fields
consists of two Chern-Simon actions of opposite level $(k,-k)$, while the ones for scalars and fermions  take the standard form in
terms of the usual  covariant derivatives. To ensure super-conformality the action is also endowed with a suitable  sextic scalar potential 
and $\psi^2 C^2$ Yukawa type interactions explicitly spelled
out in~\cite{ABJM, Benna:2008zy}.

\subsection{The 1/2 BPS  line in ABJM}
The construction of supersymmetric Wilson loops in ABJ(M) theory ~\cite{Drukker:2009hy, Cardinali:2012ru} is notably more intricate than in the four-dimensional relative $\mathcal{N}=4$ SYM~\cite{Maldacena:1998im,Rey:1998ik,Zarembo:2002an,Drukker:2007qr}. For instance, when exploring the dynamics of the 1/2 BPS heavy massive particles obtained via the Higgsing procedure~\cite{Lee:2010hk}, one discovers that they are coupled not only to bosons (as occurs in $D=4$) but to fermions as well. Then the low-energy theory of these particles turns out to possess a $U(N|N)$ supergauge invariance instead of the {\it smaller} but  {\it expected} $U(N)\times \hat U(N)$ gauge symmetry. Therefore the (locally) 1/2-BPS Wilson loop operator  must be realized as the holonomy  of a  superconnection $\mathcal{L}(t)$ living in $u(N|N)$ ~\cite{Drukker:2009hy, Lee:2010hk, Cardinali:2012ru}:
\begin{equation}\label{WL}
 \mathcal{W}=\Str\left[P\exp\left(-i\oint dt \mathcal{L}(t) \right) \mathcal{T}\right].
\end{equation}
This superconnection can be written in terms of the ABJ(M) fields and  reads
\begin{equation}\label{superconnection}
\mathcal{L}=\begin{pmatrix} A_\mu \dot{x}^\mu-\frac{2\pi i}{k} |\dot x| M_J{}^I C_I \bar C^J & -i\sqrt{\frac{2\pi}{k}}|\dot{x}|\eta_I \bar \psi^I\\
             -i\sqrt{\frac{2\pi}{k}}|\dot{x}| \psi_I \bar \eta^I & \hat A_\mu \dot x^\mu-\frac{2\pi i}{k} |\dot x| \hat{M}^I{}_J  \bar C^J C_I
            \end{pmatrix}
\end{equation}
Here $x^{\mu}(t)$ parametrizes the contour while the matrices $M_J{}^I(t)$ and $\hat M^I{}_J(t)$ and the spinors $\eta_I(t)$ and $\bar \eta^I(t)$ are local couplings,  determined in terms of the circuit $x^{\mu}(t)$ by the requirement of preserving some of the supercharges. 
The invariance for this type of loop operators does not follow  from imposing  $\d_\text{susy}\mathcal{L}=0$  as usual, but   the weaker condition
\begin{equation}\label{susyvarL}
\d_\text{susy}\mathcal{L}=\mathcal{D}_{t} \mathcal{G}=\pa_{t} \mathcal{G}+i[\mathcal{L},\mathcal{G}] 
\end{equation}
where $\mathcal{G}$ is a $u(N|N)$ supermatrix. Namely the action of supersymmetry on the connection $\mathcal{L}(t)$ can be cast as an infinitesimal supergauge transformation of   $U(N|N)$ \cite{Drukker:2009hy,Lee:2010hk,Cardinali:2012ru}. This directly implies a vanishing variation for the (super)traced Wilson loop, provided that 
 $\mathcal{G}$ is periodic along the contour.  When $\mathcal{G}$ is not exactly periodic, one can correct the  lack of periodicity   either by inserting a twist supermatrix $\mathcal{T}$ in the supertrace  (see  eq. \eqref{WL}) ~\cite{Cardinali:2012ru} or by adding to $\mathcal{L}(t)$ a background connection living on the contour  ~\cite{Drukker:2019bev}. The explicit form of either  $\mathcal{T}$  or the background connection is not relevant for the subsequent 
analysis.

Below we shall focus on the straight line case.   The line is located at  $x^2=x^3=0$ and the couplings  in \eqref{superconnection} are
given by
\be
\renewcommand*{\arraystretch}{0.8}
\!\!\!\!\!\! M_J{}^I =\hat{M}^I{}_J  =\left(~\begin{array}{cccc}
                    \!\!\!-1 	&0	&0	&0\\
                   0	&1		&0	&0\\
                    0		&0				&1	&0\\
                    0		&0				&0	&1
                   \end{array}\right) \, , \quad \mathrm{and}\quad
\eta_I^\a=
                  \left( \begin{array}{c}
                    1\\
                    0\\
                    0\\
                    0
                   \end{array}\right)_{\!\! I}\!\!
                   \eta^+, \ \ 
\bar\eta^I_\a=i
                   \begin{pmatrix}
                    1&
                    0&
                    0&
                    0
                   \end{pmatrix}^I\!\!
               \bar\eta_+
\ee
 where $\eta^+=\bar\eta_+^T=\begin{pmatrix}1 & 1\end{pmatrix}$.  These couplings break the original symmetry  OSp$(6|4)$ to
 SU$(1,1|3)$: in particular the original bosonic subsector of the supergroup containing the Euclidean conformal group in three-dimensions Sp$(4)\simeq$ SO$(1, 4)$ and the R-symmetry group SO$(6)\simeq$ SU$(4)$ reduces to  SU$(1,1) \times$  SU$(3)_R \times$ U$(1)_{J_0}$. The first factor SU$(1,1)\simeq$ SO$(2,1)$ is simply the conformal algebra in one dimension,   SU$(3)_R$ is the residual R-symmetry group  and the U$(1)_{J_0}$ factor is a recombination of the rotations around the line and a broken R-symmetry diagonal generator. The preserved generators are given in appendix \ref{algebra}.  The structure of the residual supergroup implies that we have a defect SCFT$_1$ living along the straight-line. Its operators  are characterized by a set of four quantum numbers $[\Delta, j_0, j_1, j_2]$ associated to the four Cartan generators of the above bosonic subalgebra\footnote{$\Delta$ is the conformal dimensions, $j_0$ is the  U$(1)_{J_0}$ charge and $j_1$ and $j_2$ are the SU$(3)_R$ labels.}. The structure of short and long multiplets representing this subalgebra has been studied thoroughly  in~\cite{defectABJM} and we review it in Appendix~\ref{representations}. 
Of particular relevance within them is the displacement supermultiplet which we review in the next subsection.

In the following,  we will also find  convenient to accomodate the original scalar and fermionic fields of ABJ(M) theory according  the new R-symmetry pattern
\begin{align}
 C_I&=(Z, Y_a) &  \bar C^I&=(\bar Z, \bar Y^a)\\
 \psi_I^\pm&=(\psi^{\pm},\chi_a^{\pm}) &\bar \psi^I_\pm&=(\bar\psi_{\pm},\bar\chi^a_{\pm})\label{fermions}
\end{align}
where $Y^a$ ($\bar Y_a$) and $\chi^{a\pm}$ ($\bar\chi_{a\pm}$) change in the $\mathbf{3}$ ($\mathbf{\bar 3}$) of SU$(3)$, whereas $Z$ and $\psi^{\pm}$ $(\bar \psi_{\pm})$ are singlet.  Moreover,   we have expressed them in a basis of eigenvectors of $\gamma_1=\sigma_1$, e.g. \begin{equation}
 \psi_+=(\psi_1+\psi_2) \qquad \psi_-=(\psi_1-\psi_2)
\end{equation}
with the rules $\psi^-=-\psi_+$ and $\psi^+=\psi_-$.
The two gauge fields and consequently the covariant derivative can be instead split according to the new spacetime symmetry pattern, namely
\begin{equation}\label{gaugecomps}
 A_{\mu}=(A_1,A=A_2-i A_3,\bar A=A_2+i A_3)\qquad \hat A_{\mu}=(\hat A_1,\hat A=\hat A_2-i \hat A_3,\hat{ \bar{ A}}=\hat A_2+i \hat A_3)
\end{equation}
and $D_\mu=(D_1,D=D_2-i D_3,\bar D=D_2+i D_3) .$  In terms of  these new fields the super-connection \eqref{superconnection} for the case of the straight line takes the following form 
        \begin{align}\label{straightlineconna}
 \mathcal{L}(t)&=\begin{pmatrix}  A_1  & 0\\
             0 & \hat A_1 
            \end{pmatrix}+\frac{2\pi i}{k}\begin{pmatrix}  Z \bar Z-Y_a \bar Y^a  & 0\\
             0 & \bar Z Z -\bar Y^a Y_a 
            \end{pmatrix} + \sqrt{\frac{2\pi}{ k}} \begin{pmatrix} 0  & - i\bar \psi_+ \\
            \psi^+ & 0
            \end{pmatrix}.
\end{align}

\bigskip

\subsection{The displacement supermultiplet}\label{dispmult}
An infinitesimal variation of the Wilson line \eqref{WL} translates into an operator insertion according to the identity:
\begin{equation}\label{insert}
 \frac{\braket{(\delta\mathcal{W}) \dots}}{\braket{\mathcal{W}}}=-i\int_{\mathcal{C}} dt\,\langle \delta \mathcal{L}(t)\dots \rangle_{\mathcal{W}}.
\end{equation}
where on the l.h.s. we are considering an arbitrary correlator with the deformed Wilson line, while on the r.h.s. we are using the definition \eqref{npoint}. Notice that $\delta$ could be any symmetry generator that is not preserved by the Wilson line. When $\delta$ is the action of a broken fermionic  or bosonic charge inside the parent superalgebra, the resulting operators  can be related to  an element of the 
super-multiplet of the 
displacement operator. Consider, for instance, the  six broken R-symmetry generators $J_1{}^a\equiv\mathsf{J}^a$ and $J_a{}^1\equiv\bar{\mathsf{J}}_a$ (see appendix \ref{algebra}). Their action will yield  six bosonic defect operators by the symbolic action
\begin{align}\label{O}
 [\mathsf{J}^a, \mathcal{W}]&\equiv i\delta_{\mathsf{J}^a} \mathcal{W}= \int dt~ \mathcal{W}[\mathbb{O}^a(t)]  & [\bar{\mathsf{J}}_a, \mathcal{W}]&\equiv i\delta_{\bar{\mathsf{J}}_a} \mathcal{W}= \int dt~ \mathcal{W}[\bar{\mathbb{O}}_a(t)] 
 \end{align}
where we indicate by $\mathcal{W}[\mathbb{O}^a(t)]$ an operator that is inserted on the Wilson line according to the definition \eqref{WOOO}. The equations \eqref{O} are Ward identities and they need to be thought as inserted in some correlation function. The defect operator $\mathbb{O}^a(t)$ has  conformal dimension  $\Delta=1$, since  the line defect is dimensionless and the dilatations commute with $\mathsf{J}^a$. The $U(1)_{J_0}$ charge is $2$ and it can be read from the commutator $[J_0,\mathsf{J}^a]=-2[ J_1{}^1,\mathsf{J}^a]=2 \mathsf{J}^a$. Finally, it transforms in the fundamental representation of SU$(3)$. Namely this operator is characterized by the following set of four 
quantum numbers $[1,2,1,0].$  Similarly for $\bar{\mathbb{O}}_a(t)$  we have $[1,-2,0,1].$

Next to the  $R-$symmetry broken generators, we also have  six broken supercharges given by $Q^{1a}_-\equiv i \bar{\mathsf{Q}}^a$ and $Q^{ab}_{{+}}=\e^{abc}\mathsf{Q}_c$ and thus we can define six fermionic defect operators:
\begin{align}
\label{La}
 [\mathsf{Q}_a, \mathcal{W}]&=\int dt ~\mathcal{W}
[\mathbb{\L}_a(t) ] &  [\bar{\mathsf{Q}}^a, \mathcal{W}]&=\int dt ~\mathcal{W}
[ \bar{\mathbb{\L}}^a(t) ]
 \end{align}
The  quantum numbers  of these operators are again fixed by the quantum numbers of the line defect and the commutation relations of the broken charges with the preserved ones. We find  $[\frac32,\frac52,0,1]$ for $\mathbb{\L}_a(t)$   and $[\frac32,-\frac52,1,0]$ for
$ \bar{\mathbb{\L}}^a(t)$. 

 Finally,  we  can consider the two  broken translations $\mathsf{P}$ and $\bar{\mathsf{P}}$ in the directions orthogonal to the defect. They define the so-called  displacement operators
\begin{align}\label{D}
  [\mathsf{P}, \mathcal{W}]&=\int dt  \mathcal{W}[\mathbb{D}(t)] & [\bar{\mathsf{P}}, \mathcal{W}]&=\int dt  \mathcal{W}[\bar{\mathbb{D}}(t)]
\end{align}
with charges $[2,3,0,0]$ and $[2,-3,0,0]$ respectively. 

The above construction in general does not uniquely determine the above set of operators. Since they are defined as objects inserted in the defect and we integrate over the position along the loop, we can add  a total derivative with respect to $t$  to the integrand  without altering the result.

The set of defect operators obtained through the action of  the broken charges organizes itself as a super-multiplet. The action of the preserved supercharges $
 Q^a=Q^{1a}_{+}$ and  $\bar Q_a=i\,\frac12 \epsilon_{abc} Q_{-}^{bc} $
 on them is essentially dictated by the commutation relations between  broken and preserved charges. For instance, to compute $[Q^a,\mathbb{O}^b]$ one has simply to consider the commutator $[Q^a,\mathsf{J}^b]$ acting on $\mathcal{W}$  and we get
\begin{align}
 [Q^a,\mathsf{J}^b]&=Q^{ab}_+=\e^{abc}\mathsf{Q}_c  &\Rightarrow& & [Q^a,\mathbb{O}^b]&=\e^{abc}\mathbb{\L}_c.
\end{align}
From the commutation relation of  $Q_a$ with the broken supercharges and translation  we immediately find
\be
\{Q^a,\mathbb{\L}_b\}&=-2 \d^a_b \mathbb{D}\qquad \text{and}\qquad [Q^a,\mathbb{D}]=0. 
\ee
Fixing the action of $\bar Q_a$ on these defect operators requires more attention. In fact the naive application of the above procedure
would give zero since $\bar Q_a$ commutes with all the broken charges. But this is  inconsistent 
with  $\{Q^a,\bar Q_b\}=2 \d^a_b P$ where $P$ generates the translations along the line.  However, as stressed above, any result obtained from \eqref{O}, \eqref{La} and  \eqref{D} is
defined up to  a derivative with respect to $t$, which yields zero when integrated.  To fix the form of  these derivatives in the commutation relations is more convenient to use the super-Jacobi identities.  For instance, we can determine $[\bar Q_c,\mathbb{D}]$ as follows\footnote{We choose to represent $P$ as $-\partial_t$ and then $[P,\Phi]=\partial_t\Phi$, see the discussion in \cite{Fortin:2011nq}.}
\begin{align}
0=&\{\bar Q_c,[Q^a,\mathbb{D}]\}-\{Q^a,[\mathbb{D},\bar Q_c]\}+ [\mathbb{D},\{\bar Q_c,Q^a\}]=\{Q^a,[\bar Q_c,\mathbb{D}]\}-2\delta^a_c[P,\mathbb{D}]=\nonumber\\=&\{Q^a,[\bar Q_c,\mathbb{D}]\}-2\delta^a_c\partial_t\mathbb{D}=
\{Q^a,[\bar Q_c,\mathbb{D}]+\partial_t \mathbb{\Lambda}_c\}\qquad\Rightarrow\qquad[\bar Q_c,\mathbb{D}]=-\partial_t \mathbb{\Lambda}_c.
\end{align}
where we assumed, as usual, that the action of the translation $P$ is realized by  $-\partial_t$ to be consistent with \eqref{generators-diff}.
Similarly we can show that  $ \{\bar{Q}_a,\mathbb{\L}_b\}=-2 \e_{abc}\pa_{t}\mathbb{O}^c$. Finally we have to consider the action,
since we do not have an operator of lower dimension, it would be natural to set $[\bar Q_c,\mathbb{O}^b]=0$: this choice is, however, inconsistent
with the super-Jacobi identity:
\begin{align}
\label{SJac}
0=&\{\bar Q_c,[Q^a, \mathbb{O}^b]\}-\{Q^a,[\mathbb{O}^b,\bar Q_c]\}+ [\mathbb{O}^b,\{\bar Q_c,Q^a\}]=-2\delta_c^{b}\partial_t\mathbb{O}^{a}
+\{Q^a,[\bar Q_c,\mathbb{O}^b]\}.
\end{align}
The consistency of \eqref{SJac}  suggests the existence of an additional fermionic operator $\mathbb{F}$, that is a singlet under SU$(3)$,
which obeys the anticommutation relation:
\begin{align}
 \{Q^a,\mathbb{F}\}&=\mathbb{O}^a, 
\end{align}
which in turn implies $[\bar Q_c,\mathbb{O}^b]=2 \delta^b_c \partial_t \mathbb{F}$.  The Jacobi identity for this new field immediately shows that   $ \{\bar Q_a,\mathbb{F}\}$ can be consistently chosen to vanish. Furthermore, $\mathbb{F}$ has the correct quantum numbers to be the superprimary of the \emph{chiral} multiplet $\bar{\mathcal{B}}^{\frac12}_{\frac32,0,0}$ (see Appendix \ref{representations}) with the structure 
 \begin{align} \bar{\mathcal{B}}^{\frac12}_{\frac32,0,0}: &\nonumber\\ &
\begin{tikzpicture}
 \draw[->]  (-5,5)--(-4.7,4.7);
  \node[above] at (-5.5,5) {$[\frac12,\frac32,0,0]$};
 \draw[->]  (-4,4)--(-3.7,3.7);
  \node[above] at (-4.5,4) {$[1,2,1,0]$};
\draw[->]  (-3,3)--(-2.7,2.7);
  \node[above] at (-3.5,3) {$[\frac32,\frac52,0,1]$};
  \node[above] at (-2.5,2) {$[2,3,0,0]$};
  \end{tikzpicture}
  \label{displacement}
\end{align}
In Section~\ref{sec:weakcoupling}  we will illustrate how   the supermultiplet  is realized a in terms of  the fundamental fields of  ABJM theory. 
An analogous actions of the supercharges can be derived for the conjugated operators leading to the barred version of the commutation relations in table \ref{tab:susy-displacement} up to changing the signs of the terms involving an epsilon tensor.

\begin{table}[h]
\caption*{\sc Summary of supersymmetry transformations}
\centering
{\renewcommand{\arraystretch}{1.5}%
\begin{tabular}{ l c c l }
\hline
$\{Q^a,\mathbb{F}\}=\mathbb{O}^a $ &  &  &$ \{\bar Q_a,\mathbb{F}\}=0$ \\ 
$[Q^a,\mathbb{O}^b]=\e^{abc}\mathbb{\L}_c $ &  &  &$[\bar Q_a,\mathbb{O}^b]=2 \delta^b_a \partial_t \mathbb{F}$ \\  
$\{Q^a,\mathbb{\L}_b\}=-2 \d^a_b \mathbb{D}$ &  &  &$ \{\bar{Q}_a,\mathbb{\L}_b\}=-2 \e_{abc}\pa_{t}\mathbb{O}^c$ \\
$[Q^a,\mathbb{D}]=0$& & & $[\bar Q_c,\mathbb{D}]=-\partial_t \mathbb{\Lambda}_c$\\
\hline
\end{tabular}}
\medskip
\caption{\small Summary of the supersymmetry transformations of the displacement supermultiplet of a 1/2 BPS line-defect in  $\mathcal{N}=6$ supersymmetric theories in $D=3$.}
\label{tab:susy-displacement}
\end{table}

 The displacement supermultiplet \eqref{displacement} appearing in the defect CFT$_1$ living along the Wilson line
 should match  the string transverse excitations via AdS/CFT. 
As reviewed in the Introduction and below in Section~\ref{sec:sigmamodel}, the expansion of the Green-Schwarz string action   around the minimal surface solution results in a multiplet of fluctuations transverse to the string~\cite{Forini:2012bb, Aguilera-Damia}  whose components match precisely the  quantum numbers of the operators in~\eqref{displacement},  once the two standard relations between AdS$_{2}$  masses and the corresponding CFT$_1$ operator dimensions  are taken into account, $m^2=\Delta (\Delta-1)$ for scalars and $\Delta=\frac{1}{2}+|m|$  for spinors~\cite{Henningson:1998cd,Mueck:1998iz}.  In particular,  in the scalar fluctuation sector one finds one  massive  (with $m^2=2$)  complex scalar field $X$ in AdS$_4$, corresponding to the $\Delta=2$ displacement operator $\mathbb{D}$, and  three  massless  \emph{complex} scalar fluctuations $w^a, ~a=1,2,3$ in $\cp^3$, corresponding to the $\Delta=1$ operators $\mathbb{O}^a,  ~a=1,2,3$. In the fermionic sector, there are two massless fermions which should correspond  to the $\Delta=\frac{1}{2}$ fermionic superprimary $\mathbb{F}$ of the multiplet and its conjugate, as well as six massive fermions (of which three with mass $m_F=1$ and three with $m_F=-1$) corresponding to the $\Delta=\frac{3}{2}$ fermionic operator $\mathbb\Lambda_a$ and its conjugate. 

\subsection{Weak coupling representation  in terms of supermatrices}
\label{sec:weakcoupling}

The goal of this  subsection is to provide an explicit realization of the ABJM displacement supermultiplet  in terms of the fundamental fields of the underlying theory. Since these fields live along the Wilson lines and in most cases they are obtained by varying the  superconnection \eqref{superconnection}, they naturally possess the structure of a supermatrix. 
The lowest component of the multiplet $\mathbb{F}$ has quantum numbers $[\frac{1}{2},\frac{3}{2},0,0]$ and thus we expect it to be a merely fermionic object. On the other hand,  the only field or combination of elementary fields  with these quantum numbers, which  can appear in the
entries of the super-matrix, is the bosonic complex scalar $Z$. Therefore we shall write the following  ansatz for $\mathbb{F}$, 
\be\label{Fmatrix}
\mathbb{F}=i\sqrt{\frac{2\pi}{k}}\,\tilde\epsilon\begin{pmatrix} 0 & Z\\0 & 0\end{pmatrix}\,,
\ee
where $\tilde\epsilon$ is a fermionic parameter endowing $\mathbb{F}$ with its anticommuting nature~\footnote{We remark that $\tilde\epsilon$ is just a bookkeeping device to keep memory of the Gra\ss mann nature of the operator insertions when constructing their explicit representation in terms of super matrices, and it appears in the definitions \eqref{Fmatrix}, \eqref{Omatrix}, \eqref{Lambdamatrix}, \eqref{Dmatrix} of $\mathbb{F}, \mathbb{O}, \mathbb{\Lambda}, \mathbb{D}$  below. In any explicit gauge theory computation of four-point functions the presence of such fermionic parameters (each of the fields in the correlator shold have a different one) will result in an overall factor, which should be dropped when comparing the final result with any alternative calculation. In particular, there is no $\tilde\epsilon$ dependence  in the correlators of the quantum fluctuations of the dual fundamental string evaluated in Section~\ref{sec:sigmamodel}. The fermionic/bosonic nature of these fluctuations is indeed already explicit in the corresponding fluctuation Lagrangian.}. To construct the next element of the multiplet we have to act with $\theta_a\,Q^a$ on the operator inserted in the Wilson line, where $\theta_a$ is a Gra\ss mann-odd parameter that we can identify with one of the fermionic coordinates of  the superspace constructed in sec. \ref{sec:superspace}. In the four-dimensional case, the only contribution would come from the action of $\theta_a\,Q^a$ directly on the operator,  
since the connection and thus the (open) Wilson line are both invariant.  This is not the case for operators inserted in  the fermionic Wilson loop of ABJM  theory. In fact there is an additional contribution coming from the variation of the Wilson line.  However 
this amounts to ``covariantize''  the usual action of the supersymmetry on the operator as follows
\begin{align}
\label{deltacov}
\delta_{\text{cov}}\, \sbullet[0.75] &= \theta_a \,Q^a  \sbullet[0.75] +2 \, \sqrt{\frac{2\pi}{k}}\,\left[\begin{pmatrix}0&0\\ \,\theta_a\,\bar Y^a&0
\end{pmatrix}\,, \sbullet[0.75]  \right]\,, \\
\bar\delta_{\text{cov}} \, \sbullet[0.75]&=\bar \theta^a \,\bar{Q}_a \sbullet[0.75]+ 
2 \, \sqrt{\frac{2\pi}{k}}\,\left[\begin{pmatrix}0&\bar\theta^a Y_a\\ 0&0
\end{pmatrix}\,,\sbullet[0.75]\right]\,,\label{deltacovbar}
\end{align}
where $\theta_a$ and $\bar \theta^a$ are Gra\ss mann odd parameters associated to supertranslations. The commutators in \eqref{deltacov} and \eqref{deltacovbar} compensates the super-gauge transformation of the Wilson line induced by the supersymmetry transformation \eqref{susyvarL} with the matrix $\mathcal{G}$ given by
\begin{align}
 \mathcal{G}=2 \, \sqrt{\frac{2\pi}{k}}\begin{pmatrix}0&\bar\theta^a Y_a\\ \,\theta_a\,\bar Y^a&0
\end{pmatrix} \, .
\end{align}

The action of $\delta_{\text{cov}}$ on   $\mathbb{F}$ defined in \eqref{Fmatrix} is quite straightforward to evaluate once   we use the transformations in Appendix~\ref{app:susytransf}. The final result is
\be\label{Omatrix}
\delta_{\text{cov}}\mathbb{F}=\, i\sqrt{\frac{2\pi}{k}} \theta_a\,\tilde\epsilon
\begin{pmatrix}
2\sqrt{\frac{2\pi}{k}}\,Z\,\bar Y^a & - \bar\chi^a_+\\ 
 0&2\sqrt{\frac{2\pi}{k}}\,\bar Y^a\,Z 
\end{pmatrix}\equiv 
\theta_a\,\mathbb{O}^a\,.
\ee
namely we have obtained the second component of our supermultiplet~\eqref{displacement}, the one associated to the $R$-symmetry breaking. An identical expression can be obtained by exploiting that the fields appearing in the 
super-connection \eqref{straightlineconna}  under the action of $\mathsf{J}^a$ transforms as follows
\begin{equation}
{\delta_{\mathsf{J}^a}(Z,{Y_b})=(0, i \delta^a_b Z),\qquad
{\delta_{\mathsf{J}^a}(\bar Z,\bar {Y}^b})=(-i\bar {Y}^a},0),\qquad {\delta_{\mathsf{J}^a}}\psi^+=0,\qquad  {\delta_{\mathsf{J}^a}}\bar\psi_+=-i\bar\chi^a_+,
\end{equation}
Using \eqref{O} one can apply the action of these broken generators on the superconnection \eqref{straightlineconna} recovering the same result for $\mathbb{O}^a$.
Similarly, the form of $\bar{\mathbb{O}}_a(t)$ can be obtained  by looking at the explicit action of $\bar{\mathsf{J}}_a$ on the fields. 
Applying once more $\delta_{\text{cov}}$, we reach the third component  
\begin{align} \nonumber
\delta_{\text{cov}}\,\big(\,\theta_a\,\mathbb{O}^a\,\big)&=2i\sqrt{\frac{2\pi}{k}}\, \epsilon^{nrs}\,\theta_r \theta_s\,\tilde\epsilon
\begin{pmatrix}
\sqrt{\frac{2\pi}{k}}\,(\epsilon_{abn} \,\bar\chi^a_+ \,\bar Y^b - Z\,\chi^-_n)
& - D\, Y_n \\ 
\frac{4\pi}{k}\, \epsilon_{abn} \bar Y^a \,Z \bar Y^b&\sqrt{\frac{2\pi}{k}} \,(\epsilon_{abn} \,\bar Y^b\,\bar\chi^a_+ -\chi^-_n \,Z\,)
\end{pmatrix}
\\ \label{Lambdamatrix}
 & \equiv-
\epsilon^{nrs}\,\theta_r \,\theta_s\,\mathbb{\Lambda}_n\,.
\end{align}
Also in this case, the expression \eqref{Lambdamatrix} for $\mathbb{\Lambda}_n$ perfectly agrees with the one obtained by acting with the broken generator on the superconnection \eqref{straightlineconna} according to \eqref{La}. 

Finally,  applying again $\delta_{\text{cov}}$ to \eqref{Lambdamatrix} we obtain the top component $\mathbb{D}$ of the supermultiplet \eqref{displacement}, namely the displacement operator
\begin{align} \nonumber
\!\!\!\!\!
\delta_{\text{cov}} \big( - \epsilon^{nrs}& \theta_r  \theta_s  \mathbb{\Lambda}_n \big)
\!=\! 2   \sqrt{\frac{2\pi}{k}} \epsilon^{nrs}\theta_n \theta_r  \theta_s\,\tilde\epsilon\! ~~\times\\ &\times\begin{pmatrix}\!
 \sqrt{\frac{2\pi}{k}}\, i(2Z\, D\bar Z- 2D Y_a\bar Y^a +\bar\chi^a_+\chi^-_a) & -i \,D\bar\psi_+
 \\ \nonumber
D\psi_+ -i\,\widetilde{\mathcal{D}}_t\psi^-
 & 
 \sqrt{\frac{2\pi}{k}}\, i(2 D\bar Z\,Z -2 \bar Y^a\, D Y_a -\chi^-_a\,\bar\chi^a_+) \!
 \end{pmatrix}\equiv
\\ \label{Dmatrix}
 & \equiv 2
\epsilon^{nrs}\,\theta_n \,\theta_r  \,\theta_s \,\mathbb{D}\,.
\end{align}
where $\widetilde{\mathcal{D}}_t$ is the covariant derivative constructed only with the bosonic part of the super-connection.
 The $\mathbb{D}$ insertion can be also constructed out of its abstract definition~\eqref{D} by acting with a broken translation on the superconnection, as was done in equation (5.18) of~\cite{defectABJM}. In this case, the two expressions are not identical and the difference between the two definitions is proportional to the matrix
\be
\begin{pmatrix}
\sqrt{\frac{2\pi}{k}}\bar\psi_+ \,\psi^- &  0
 \\ 
\,\widetilde{\mathcal{D}}_t \psi^-  & - \sqrt{\frac{2\pi}{k}}\psi^-\,\bar\psi_+
 \end{pmatrix}
 {  \propto}\, \mathcal{D}_t
 \begin{pmatrix}
0 &  0
 \\ 
\psi^-  &0
 \end{pmatrix}\,,
\ee
This difference is compatible with the construction~\eqref{D}, which is blind to total derivatives.
For the case of Wilson lines insertions, total derivatives are implemented~\cite{defectABJM} as the following modification for an operator $\mathcal{O}(t)$ inserted into the loop
\begin{align}
\int dt ~\mathcal{W}
[\mathcal{O}(t)]=&\int dt ~[\mathcal{W}
[\mathcal{O}(t)]+\partial_t(\mathcal{W}
[ \Sigma(t) ] )]=\nonumber\\=&\int dt ~(\mathcal{W}
[\mathcal{O}(t)]+\mathcal{W}
[\mathcal{D}_t \Sigma(t) ] )=
\int dt ~\mathcal{W}
[\mathcal{O}(t)+\mathcal{D}_t \Sigma(t) ] \,,
\end{align}
where $\mathcal{D}_t$ is the covariant derivative defined in~\eqref{susyvarL}. 
 We then conclude that our supermatrix construction perfectly agrees with the abstract structure outlined in Section \ref{dispmult}.


\section{Chiral correlation functions in superspace}
 \label{sec:superspace}

\subsection{Algebra}
 
The algebra preserved by the 1/2 BPS Wilson line is the 
one-dimensional $\mathcal{N}=6$ superconformal algebra, or $su(1,1|3)$, generated by $\{D,P,K,{R_a}^b, J ;Q^a,\overline{Q}_a,S^a,\overline{S}_a\}$, $a=1,2,3$,  where the identifications with the  $osp(6|4)$ generators are spelled out in Appendix~\ref{app:algebra-superconf}. 
The bosonic generators  $\{D,P,K\}$ are those  of the one-dimensional conformal algebra $su(1,1)\sim so(2,1)$, together with the SU$(3)$ traceless generators ${R_a}^b$ ($a,b=1,2,3$) and the additional $u(1)$ R-symmetry current algebra generator $J_0$.  Their commutation  relations read
\begin{align}
 [P,K]&=-2 D\,,& ~~ [D,P]&=P\,, &  [D,K] &=-K & [{R_{a}}^{b},{R_{c}}^{d}]&=\d_{a}^{d} {R_{c}}^{b} -\d^{b}_{c} {R_{a}}^{d}\,,
 \end{align}
 and $J_0$ commutes with all of them.
The anticommutation relations for the fermionic generators $Q^a$, $\overline{Q}_a$ and the corresponding superconformal charges  $S^a, ~\overline{S}_a$, $a=1,2,3$ are
\begin{align}
 \{Q^a,\bar Q_b\}&=2 \d^a_b P\,, &  \{S^a,\bar S_b\}&=2\d^a_b K \\  
 \{Q^a,\bar S_b\}&= 2\d_b^a ( D+\tfrac13 J_0)-2 {R_b}^a & \{\bar Q_a, S^b\}&= 2\d_b^a(D-\tfrac13 J_0)+2 {R_a}^b \,,
 \label{anticommQS}
\end{align}
and the non-vanishing mixed commutators read \small
\begin{align}
 [D,Q^a]&=\frac12 Q^a & [D,\overline{Q}_a]&=\frac12 \overline{Q}_a &  [K,Q^a]&=S^a & [K,\overline{Q}_a]&= \overline{S}_a\\
 [D,S^a]&=-\frac12 S^a & [D,\overline{S}_a]&=-\frac12 \overline{S}_a &  [P,S^a]&=-Q^a & [P,\overline{S}_a]&=- \overline{Q}_a\\
 [{R_a}^b,Q^c]&=\d_a^c Q^b-\tfrac13 \d_a^b Q^c & [{R_a}^b, \overline{Q}_c]&=-\d_c^b \overline{Q}_a+\tfrac13 \d_a^b \overline{Q}_c & [J_0,Q^a]&=\tfrac12 Q^a & [J_0,\overline{Q}_a]&=-\tfrac12 \overline{Q}_a\\
 [{R_a}^b,S^c]&=\d_a^c S^b-\tfrac13 \d_a^b S^c & [{R_a}^b,\overline{S}_c]&=-\d_c^b \overline{S}_a+\tfrac13 \d_a^b \overline{S}_c & [J_0,S^a]&=\tfrac12 S^a & [J_0,\overline{S}_a]&=-\tfrac12 \overline{S}_a
\end{align}\normalsize
The quadratic Casimir is
\be
\label{Casimir}
{\bf C}^{(2)}=D^2-\frac{1}{2}\,\{K,P\}+\frac{1}{3}J_0^2-\frac{1}{2}{R_a}^b\,{R_b}^a+\frac{1}{4}\,[\overline{S}_a,Q^a]+\frac{1}{4}\,[S^a,\overline{Q}_a]~,
\ee
and, when acting on a highest weight state $[\Delta,j_0,j_1,j_2]$  of $su(1,1|3)$, it has eigenvalue
\be\label{casimireigenv}
{\bf c_2} =   \Delta (\Delta+2)+\frac{j_0^2}{3}\,-\frac{1}{3} (j_1^2 + j_1 j_2 + j_2^2).
\ee
To compute the superconformal blocks we need to represent the algebra above as an action on superconformal primaries. For this purpose we introduce $(t, \theta_a, \bar\theta^a)$ as superspace coordinates, where $t$ is the coordinate along the Wilson line and $\theta_a$ and $\bar\theta^a$ are Gra\ss mann variables~\footnote{The natural superspace to study correlation functions in the bulk has been introduced in~\cite{Liendo:2015cgi}, and the setup adopted here could also be seen as the reduction of the one considered there.}.        We can then write the differential action of the $\mathcal{N}=6$ generators  as
\begin{align}\label{generators-diff}
P&=-\pa_{t} \\
 D&=\textstyle{-t \pa_t-\frac12 \theta_a \pa^{a} -\frac12 \bar \theta^a \bar \pa_a - \Delta}\\
 K&=-t^2 \pa_{t}-(t+\theta\bar\theta) \theta_a \pa^{a}-(t-\theta \bar\theta) \bar \theta^a \bar \pa_a-(\theta \bar \theta)^2 \pa_{t} -2\,t\, \Delta+\frac23 j_0\,\theta\bar \theta\\
 Q^a&=\pa^{a}-\bar \theta^a \pa_t\\
 \bar Q_a&=\bar \pa_{a}- \theta_a \pa_t\\
 S^a&=(t+\theta \bar \theta)\pa^a-(t-\theta \bar \theta) \bar \theta^a \pa_{t}-2\bar \theta^a \bar \theta^b \bar \pa_b-2( \Delta +\frac13 j_0 ) \bar\theta^a\\
 \bar S_a&=(t-\theta \bar \theta)\bar \pa_a-(t+\theta \bar \theta)  \theta_a \pa_{t}-2 \theta_a \theta_b \pa^b-2( \Delta -\frac13 j_0 )\theta_a\\
 J_0&=-\frac12 \theta_a \pa^a+\frac12 \bar \theta^a \bar \pa_a+j_0\\\label{generators-diff-end}
 R_a{}^b&=-\theta_a \pa^b+\bar \theta^b \bar \pa_a+\textstyle{\frac{1}{3}}\,\delta_a^b\, (\theta_c\pa^c-\bar\theta^c\bar\pa_c)
  \end{align}
where $\pa^a=\frac{\pa}{\pa \theta_a}$, $\bar \pa_a=\frac{\pa}{\pa \bar \theta^a}$, $\theta \bar \theta =\theta_a \bar \theta^a$ and we are neglecting the SU$(3)$ charges $j_1$ and $j_2$ because we will only be  interested in neutral superfields. 
The superspace is also equipped with supercovariant derivatives
\be\label{supercovder}
 \mathsf{D}^a=\pa^{a}+\bar \theta^a \pa_t\,,\qquad\qquad  \bar{\mathsf{D}}_a= \bar{\pa}_{a}+ \theta_a \pa_t\,.
\ee
 
 
%
%

\subsection{Chiral correlators}

The representation theory analysis of the supergroup SU$(1,1|3)$ reviewed in the previous section shows that the displacement operator belongs to a chiral multiplet with R-charge $j_0=\frac32$ and dimension $\Delta=\frac12$ (the conjugate operator belongs to an antichiral multiplet with opposite R-charge). For the purpose of this section we are going to consider a chiral multiplet with arbitrary R-charge $j_0$ and dimension $\D=\frac{j_0}{3}$. A chiral superfield $\Phi_{j_0}$ should respect the chirality condition
\begin{equation}\label{chiralitycondition}
 \bar{\mathsf{D}}_a \Phi_{j_0}=0\,,\qquad 
\end{equation}
for every value of $a$.
By defining the chiral coordinate $y=t+\theta_a \bar \theta^a$, such that $\bar{\mathsf{D}}_a y = 0$, one simply has the component expansion
\begin{align}\label{chiralsuperfield}
 \Phi_{j_0}(y,\theta)=\phi(y)+\theta_a \psi^a(y)-\frac12 \theta_a \theta_b \,\e^{abc}\, \eta_c(y) +\frac13 \theta_a \theta_b \theta_c \,\e^{abc} \xi(y)\,,
\end{align} 
where the numerical coefficients of each  component are fixed by consistency between the action of the supercharges on the superfield and the commutation relations in Table~\ref{tab:susy-displacement}.  \\
Similarly, the antichiral field is expanded as
\begin{align}\label{antichiralsuperfield}
 \bar\Phi_{j_0}(y,\theta)=\bar\phi(y)+\bar\theta^a \bar\psi_a(y)+\frac12 \bar\theta^a \bar\theta^b \,\e_{abc}\, \bar\eta^c(y) -\frac13 \bar\theta^a \bar\theta^b \bar\theta^c \,\e_{abc}\, \bar\xi(y)\,.
\end{align} 
The two-point function of a chiral and antichiral superfield must be expressed in terms of the chiral distance ($\bar{\mathsf{D}}_i\braket{i\bar j}=\mathsf{D}_j\braket{i\bar j}=(Q_i+Q_j)\,\braket{i\bar j}=(\bar Q_i+\bar Q_j)\,\braket{i\bar j}=0$)
\be\label{distance}
\braket{i\bar j}=y_i-y_j-2{\theta_{a}}_i \bar {\theta^a}_j\,,
\ee  
and  reads
\be\label{2-point-superfield}
\braket{\Phi_{j_0}(y_1,\theta_1) \bar \Phi_{-j_0}(y_2,\bar \theta_2)}=\frac{c_{\Phi_{j_0}}}{\braket{1\bar 2}^{\frac{2j_0}{3}}}\,,
\ee
where $c_{\Phi_{j_0}}$ is a normalization constant that in the case of the displacement supermultiplet has a physical meaning, see Section~
\ref{sec:displmultiplet}. 

\noindent In this paper we are interested in the four-point functions  
\begin{align}\label{4-point-1}
 \!\! \braket{\Phi_{j_0}(y_1,\theta_1) \bar \Phi_{-j_0}(y_2,\bar \theta_2)\Phi_{j_0}(y_3,\theta_3) \bar \Phi_{-j_0}(y_4,\bar \theta_4)}&=
 \frac{c^2_{\Phi_{j_0}}}{\braket{1\bar 2}^{\frac{2j_0}{3}}\braket{3\bar 4}^{\frac{2j_0}{3}}}\,f (\mathcal{Z})\,, \\
 \!\! \braket{\Phi_{j_0}(y_1,\theta_1) \bar \Phi_{-j_0}(y_2,\bar \theta_2) \bar \Phi_{-j_0}(y_3,\bar \theta_3)  \Phi_{j_0}(y_4, \theta_4)}&=
- \,\frac{c^2_{\Phi_{j_0}}}{\braket{1\bar 2}^{\frac{2j_0}{3}}\braket{4\bar 3}^{\frac{2j_0}{3}}}\,h (\mathcal{X})\, , \label{4-point-2}
 \end{align}
where
\begin{align} \label{supercrossratios}
\mathcal{Z}&=\frac{\braket{1\bar 2}\braket{3 \bar 4}}{\braket{1\bar 4}\braket{3{\bar 2}}}  & \mathcal{X}&=-\frac{\braket{1\bar 2}\braket{4  \bar 3}}{\braket{1\bar 3}\braket{2 \bar 4}}
\end{align}
are the two superconformal cross ratios corresponding to the two different correlators. These two invariants are built out of the chiral distance~\eqref{distance}. In the general case one may have a set of additional superconformal invariants which are nilpotent due to their Gra\ss mann nature. For long-multiplet four-point functions,  the use of such nilpotent invariants guarantees a finite truncation in the superspace expansion  (see e.g.~\cite{Cornagliotto:2017dup}). 
However, none of these invariants is compatible with the chirality condition~\eqref{chiralitycondition}. The absence of nilpotent invariants is expected for correlators of 1/2 BPS operators, and in general for 4-point functions containing two chiral and two anti-chiral operators~\cite{Fitzpatrick}. It is important to notice that both correlators \eqref{4-point-1} and \eqref{4-point-2} are ordered such that $t_1<t_2<t_3<t_4$. In higher dimensions these two correlators would be related by crossing, but in one dimension this is not the case. To make this more concrete, let us consider the bosonic part of the cross ratios \eqref{supercrossratios}
\begin{align}\label{z-chi}
 z&=\frac{t_{12}t_{34}}{t_{14}t_{32}} & \chi&=\frac{t_{12}t_{34}}{t_{13}t_{24}}
\end{align}
where $z$ is the bosonic part of $\mathcal{Z}$ and $\chi$ is the bosonic part of $\mathcal{X}$. With our ordering we have $z<0$ and $0<\chi<1$. The two cross ratios are related by the transformation
\be\label{z-chi-rel}
z=\frac{\chi}{\chi-1}\,.
\ee
The absence of nilpotent invariants implies that the superprimary correlators
\begin{align}\label{4-point-1-primary}
 \!\! \braket{\phi(t_1) \bar \phi(t_2)\phi(t_3) \bar \phi(t_4)}&=\braket{\phi(t_1) \bar \phi(t_2)}\braket{\phi(t_3) \bar \phi(t_4)}\,f (z)\,, \\
 \!\! \braket{\phi(t_1) \bar \phi(t_2)\bar \phi(t_3) \phi(t_4)}&=\braket{\phi(t_1) \bar \phi(t_2)}\braket{\bar \phi(t_3) \phi(t_4)}\,h (\chi)\, , \label{4-point-2-primary}
 \end{align}
fully determine the four-point functions of the whole superconformal multiplet, i.e. the correlators of superconformal descendants can be obtained by the action of differential operators on $f(z)$. Of course, one is free to express the function $f(z)$ in terms of the cross-ratio $\chi$ by considering $f(\frac{\chi}{\chi-1})$. One can also take the analytic continuation of $f(z)$ for $0<z<1$ ($f(z)$ has branch cut singularities at coincident points, i.e. $z=0,1,\infty$), which would naively establish a relation between \eqref{4-point-1-primary} and \eqref{4-point-2-primary}. Nevertheless, in one dimension this is not the case and $h(\chi)$ is not the analytic continuation of $f(z)$. Still, we will see in section \ref{superblocks} that a relation between these two functions exists through their $s$-channel block expansion.
 
%

\subsection{Selection rules}
Even though we consider half-BPS multiplets as external operators, more general multiplets can be exchanged when an OPE is applied to the correlator.  
In our case,  there are  two qualitatively different OPE channels to consider, depending on whether we take the chiral-antichiral OPE $\Phi\times \bar\Phi$  or the chiral-chiral OPE $\Phi \times\Phi$. Each of them presents its own selection rules for superconformal representations appearing in these two channels, as well as corresponding superconformal blocks. We start from the chiral-antichiral channel. In~\cite{Bianchi:2018scb} all the selection rules for the 1/6 BPS defect theory were derived,  and it was found that only the identity and long multiplets can appear in the chiral-antichiral OPE for $su(1,1|1)$, a.k.a. $\mathcal{N}=2$ supersymmetry. For the $\mathcal{N}=6$ case of interest here, every pair $\{Q^a,\bar Q_a\}$ at fixed $a$ generates a $su(1,1|1)$ subgroup of $su(1,1|3)$, and a chiral multiplet of $su(1,1|3)$ is also a chiral multiplet of all the three $su(1,1|1)$ subgroups. This implies that the operators appearing in the $\Phi \times \bar \Phi$ OPE 
must belong to long multiplets of all the $su(1,1|1)$ subgroups, i.e. they must not be annihilated by any supercharge. Furthermore, three-point functions with one chiral, one antichiral and one long multiplet are non-vanishing only when the superprimaries R-charges sum to zero, namely a long multiplet enters only when its superprimary can be exchanged. We conclude that
\begin{align}
 \bar{\mathcal{B}}_{j_0} \times \mathcal{B}_{-j_0} \sim \mathcal{I} + \mathcal{A}^{\Delta}_{0,0,0}\,,
\end{align}
where $\mathcal{I}$ is the identity and we use the notation of Appendix~\ref{app:reprs} for the $su(1,1|3)$ supermultiplets. 
In this case the unitarity bound~\eqref{unitaritybounds} simply reads $\Delta\geq0$ and every positive dimension is allowed for long operators.

For the chiral-chiral channel, the situation is richer. To extract the selection rules we will borrow an argument from \cite{Poland:2010wg}. Consider the chiral superprimary operator $\phi$. Chirality gives
\begin{align}
 [\bar Q_a, \phi(t)]&=0 & [\bar S_a, \phi(t)]&=0
\end{align}
for any $a$ and any $t$. The first condition is simply the definition of chirality, whereas the second one comes from the requirement that $\phi$ is a superprimary at the origin together with the commutation relation $[P,\bar S_a]=- \bar Q_a$. These two conditions imply  that any operator $\mathcal{O}$ appearing in the $\phi \times \phi$ OPE must respect
\begin{align}
 [\bar Q_a, \mathcal{O}(t)]&=0 & [\bar S_a, \mathcal{O}(t)]&=0
\end{align}
for any $a$. It is then immediate to realize that the only superprimary operator which is allowed to appear is a chiral operator of dimension $\Delta_{\text{exc}}=\frac{2j_0}{3}$. All other multiplets will contribute with a single superdescendant (and all its conformal descendants) generated by the repeated action of $\bar{Q}_a$. Concretely, a long multiplet will contribute with the operator generated by $\bar{Q}^3 O$ (here $\bar{Q}^3=\e^{abc}\bar Q_a \bar Q_b \bar Q_c)$, where $O$ is the superprimary, and a similar story holds for other short multiplets. The complete analysis yields
\begin{align}
 \bar{\mathcal{B}}_{j_0} \times \bar{\mathcal{B}}_{j_0} \sim \bar{\mathcal{B}}_{2j_0} + \bar{\mathcal{B}}_{2j_0+\frac12,1} + \bar{\mathcal{B}}_{2j_0+1,0,1}+ \mathcal{A}^{\Delta}_{2j_0+\frac32,0,0} 
\end{align}
In particular, in the OPE of the superprimary operator $\phi$, every supermultiplet contributes with a single conformal family, whose conformal primary has quantum numbers $[\D,2j_0,0,0]$. As usual, for short multiplets the dimension is fixed in terms of $j_0$. In Table \ref{selrulestab} we summarize the schematic form of the only relevant superdescendant operators and we make explicit their conformal dimension, which can be easily obtained from the one of the associated superprimary. The dimension of the long multiplet is clearly unfixed, but it should respect the unitarity bounds \eqref{unitaritybounds}. For this specific case we find that the dimension of the superprimary must be\footnote{Here the equality is excluded because in that case the long multiplet decomposes as in Table \ref{longmultdec} and the relevant superdescendant falls back into the $\bar{\mathcal{B}}_{2j_0+1,0,1}$ multiplet.} $\D> \frac{2j_0}{3}+\frac12$, while the relevant superdescendant, obtained by acting with all the $\bar{Q}$'s, must have dimension 
\begin{equation}\label{boundexchange}
\D_{\text{exc}}^{\text{long}}> \frac{2j_0}{3}+2 \, .
\end{equation}
This bound will be very important in the following, where we are going to focus on the $j_0=\frac32$ case. In summary, the $\phi \times \phi$ spectrum admits three protected conformal primaries of dimensions $\frac{2j_0}{3}$, $\frac{2j_0}{3}+1$ and $\frac{2j_0}{3}+2$ and infinitely many operators with unprotected dimensions strictly higher than the protected ones. 
\begin{table}[htbp]
\begin{center}
\begin{tabular}{|l|c|c|}
 \hline
 Multiplet & Exc. & $\D_{\text{exc}}$  \\
 \hline
 $\bar{\mathcal{B}}_{2j_0}$ & $O$ & $\frac{2j_0}{3}$  \\
 $\bar{\mathcal{B}}_{2j_0+\frac12,1} $ & $ \bar{Q}O $ & $ \frac{2j_0}{3}+1 $ \\
 $\bar{\mathcal{B}}_{2j_0+1,0,1} $ &  $\bar{Q}^2O $ & $ \frac{2j_0}{3}+2 $ \\
 $ \mathcal{A}^{\D}_{2j_0+\frac32,0,0} $ & $ \bar{Q}^3O $ & $ \D+\frac32 $ \\
 \hline
\end{tabular}
\end{center}
\caption{The multiplets contributing to the chiral-chiral OPE with a schematic representation of the only superconformal descendant (but conformal primary) contributing to the OPE (in this table $O$ indicates the superprimary of each multiplet).}
\label{selrulestab}
\end{table} 
 
\subsection{Superblocks}\label{superblocks}

We now derive the superconformal blocks associated to the two channels. The chiral-chiral channel is the easier one. Each supermultiplet only contributes with a single conformal family, therefore one only needs to select the $sl(2)$ conformal blocks with the appropriate dimensions. Let us then start by introducing the $sl(2)$ blocks which resum the contribution of $1d$ conformal descendants~\cite{Dolan:2011dv}
\begin{align}\label{sl2block}
g_h(\chi)=\chi^h \, _2F_1(h,h;2 h;\chi)
\end{align}
Each conformal primary listed in Table \ref{selrulestab} contributes with an $sl(2)$ block with $h=\D_{\text{exc}}$. Let us consider the specific correlator of interest for this paper\footnote{The correlator \eqref{4-point-1-primary} does not admit an expansion in this channel since one cannot take point $\phi(t_1)$ close to $\phi(t_3)$.} 
\begin{align}\label{tildeh}
 \braket{\phi (t_1) \bar{\phi} (t_2) \bar{\phi} (t_3) \phi (t_4)}= \frac{C_{\Phi}^2}{t_{14}^{\frac{2j_0}{3}} t_{23}^{\frac{2j_0}{3}}} \hat{h}(\chi)
\end{align}
where, comparing with \eqref{4-point-2-primary} we defined $\hat{h}(\chi)=(\frac{1-\chi}{\chi})^{\frac{2j_0}{3}}h(\chi)$. The $\phi\times \phi$ channel corresponds to the $\chi\to1$ limit and we have
\begin{align}\label{tildehexpansion}
 \!\!\!\!
 \hat{h}(\chi)=\mathsf{c}_{\frac{2j_0}{3}} g_{\frac{2j_0}{3}}(1-\chi)+ \mathsf{c}_{\frac{2j_0}{3}+1} g_{\frac{2j_0}{3}+1}(1-\chi)+\mathsf{c}_{\frac{2j_0}{3}+2} g_{\frac{2j_0}{3}+2}(1-\chi)+\sum_{\Delta} \mathsf{c}_{\Delta} g_{\Delta}(1-\chi)
\end{align}
where the sum runs over unprotected operators with dimension $\Delta > \frac{2j_0}{3}+2$, and $\mathsf{c}_\Delta$'s are the squared moduli of the OPE coefficients (we use a different font to distinguish these coefficients from the ones that will appear in the chiral-antichiral channel).

In the chiral-antichiral channel, long multiplets with quantum numbers $[\Delta,0,0,0]$ contribute with all the allowed superdescendants, and we can decompose the four-point correlation function in terms of superconformal blocks. The latter are always expressed as a linear combination of ordinary $sl(2)$ blocks with shifted dimensions. Conformal blocks can be seen as eigenfunctions of the Casimir differential operator~\cite{Dolan:2011dv}. Analogously, superconformal blocks can be computed by  considering the differential equation generated by the action of the superconformal Casimir~\cite{Fitzpatrick}. We start by the $s$-channel OPE expansion of the superspace four-point function \eqref{4-point-1}, namely we insert a resolution of the identity between points $t_2$ and $t_3$, and we act on each term of the sum with the quadratic Casimir ${\bf C}_1+{\bf C}_2$, where ${\bf C}_i$ is the differential operator~\eqref{Casimir} acting on the supercoordinates $(t_i,\theta_{ai},\bar \theta^a_i)$. This leads to the block expansion
\be\label{superblock}
f(z)=1+\sum_\Delta\,c_\Delta\,G_\Delta(z)
\ee
where the sum runs over the dimensions  $\Delta>0$ of the superprimary operators exchanged in the chiral-antichiral channel and each conformal block satisfies the differential equation
\be\label{eigenfunction}
\big(-z^2(z-1)\partial_z^2 -z(z-3)\partial_z \big)\,G_\Delta (z)  = \Delta(\Delta+2)\,G_\Delta (z) \,,
\ee   
where $\Delta(\Delta+2)$ is the Casimir eigenvalue~\eqref{casimireigenv} with  zero R-charges. The equation above is solved by the hypergeometric function~\footnote{The appearance of a minus sign in~\eqref{superblock} is due to our use of $z$, which takes real negative values, rather than the more standard $\chi$, see~\eqref{z-chi}.}
\be\label{solsuperblock}
G_\Delta(z)=  (-z)^\Delta {}_2F_1(\Delta,\Delta,2\Delta+3; z)\,.
\ee
As a check, we find that $G_\Delta(z)$ can be decomposed in terms of a finite sum of $sl(2)$ blocks \eqref{sl2block} (here we use $\hat{g}(z)=g(\frac{z}{z-1})=(-z)^\Delta \, _2F_1(\Delta,\Delta;2 \Delta;z)$, implementing the change of variable from $\chi$ to $z$)
\begin{align}
G_\Delta(z)&=\hat{g}_\Delta(z)+ \frac{3 \Delta}{2 (2 \Delta+3)}   \hat{g}_{\Delta+1}(z)+ \frac{3 \Delta^2 (\Delta+1)}{4 (2\Delta+3)(2\Delta+1)(\Delta+2)}\hat{g}_{\Delta+2}(z) \nonumber  \\ 
&+\frac{\Delta^2 (\Delta+1)}{8 (2 \Delta+3)^2 (2 \Delta+5)}\hat{g}_{\Delta+3}(z) \label{expansion-conformal-blocks}
\end{align}

With the aim of finding a connection between the two correlators \eqref{4-point-1-primary} and \eqref{4-point-2-primary} it is important to consider the $s$-channel expansion of \eqref{4-point-2-primary}
\begin{align}\label{hexpansion}
 h(\chi)&=1+\sum_\Delta\,\tilde{c}_\Delta\,\tilde{G}_\Delta(\chi)
\end{align}
where 
\begin{align}
 \tilde{G}_\Delta(\chi)=\chi^\Delta {}_2F_1(\Delta,\Delta,2\Delta+3; \chi)\,.
\end{align}
The similarity between this block expansion and \eqref{superblock} is apparent, especially when one considers the relation between $\tilde c_\Delta$ and $c_\Delta$. Denoting by $\mathcal{O}_\Delta$ the exchanged operator of dimension $\Delta$, the expressions of $c_\Delta$ and $\tilde c_\Delta$ are given by
\begin{align}
 c_\Delta&= f_{\phi \bar \phi \mathcal{O}_\Delta} f_{\phi \bar \phi \mathcal{O}_\Delta} & \tilde c_\Delta=f_{\phi \bar \phi \mathcal{O}_\Delta}f_{\bar \phi  \phi \mathcal{O}_\Delta}
\end{align}
where $f_{\phi \bar \phi \mathcal{O}_\Delta}$ is the three-point coefficient determining $\braket{\phi \bar \phi \mathcal{O}_\Delta}$. In one-dimensional CFT the OPE coefficients depend on the signature of the permutation. In particular, despite there is no continuous group of rotation, there is a $\mathbb{Z}_2$ parity transformation $t\to -t$ (this symmetry was called S-parity in \cite{Billo:2013jda}). Operators are charged under this symmetry and one has
\begin{align}
 \braket{\mathcal{O}_1(t_1)\mathcal{O}_2(t_2)\mathcal{O}_3(t_3)}=(-1)^{T_1+T_2+T_3} \braket{\mathcal{O}_3(-t_3)\mathcal{O}_2(-t_2)\mathcal{O}_1(-t_1)}
\end{align}
where $T_1$, $T_2$ and $T_3$ are the charges of the operators under parity. For our example, if $\phi$ is a bosonic operator we have
\begin{align}
f_{\phi \bar \phi \mathcal{O}_\Delta}= (-1)^{T_{\mathcal{O}}} f_{\bar \phi  \phi \mathcal{O}_\Delta} 
\end{align}
whereas for a fermionic $\phi$
\begin{align}
 f_{\phi \bar \phi \mathcal{O}_\Delta}=(-1)^{1+T_{\mathcal{O}}} f_{\bar \phi  \phi \mathcal{O}_\Delta}
\end{align}
Therefore the two coefficients $c_\Delta$ and $\tilde c_\Delta$, for a fermionic $\phi$ are related by
\begin{align}\label{ctildec}
 c_\Delta=(-1)^{T_{\mathcal{O}}+1} \tilde c_\Delta
\end{align}
As an example, which will be useful in the following, operators of the schematic form $\mathcal{O}_n=\phi\pa_t^n\bar \phi$ have charge $T_{\mathcal{O}}=n$. In Section \ref{bootstrap} these relations will allow us to establish a precise connection between the correlators \eqref{4-point-1-primary} and \eqref{4-point-2-primary} in a perturbative expansion around the free theory result.

\subsection{The displacement superfield and its four-point function}
\label{sec:displmultiplet}

The displacement operator belongs, as shown in Appendix~\ref{dispmult},  to a chiral supermultiplet with $j_0=\frac{3}{2}$. In light of the application to string sigma-model in Section \ref{sec:sigmamodel}, we use the superspace analysis to extract the correlators of some relevant components of the displacement supermultiplet. The component expansion of the chiral and antichiral superfield (compared to \eqref{chiralsuperfield} we renamed the components for the specific case $j_0=\frac32$) reads
\begin{align}\label{dispsuperfield}
 \Phi(y,\theta)&=\mathbb{F}(y)+\theta_a \mathbb{O}^a(y)-\frac12 \theta_a \theta_b \,\e^{abc}\, \mathbb{\L}_c(y) +\frac13 \theta_a \theta_b \theta_c \,\e^{abc} \mathbb{D}(y)\,,\\
 \bar \Phi(y,\theta)&=\bar{\mathbb{F}}(y)+\bar\theta^a \bar{\mathbb{O}}_a(y)+\frac12 \bar\theta^a \bar\theta^b \,\e_{abc}\,\bar{\mathbb{\L}}^c(y) -\frac13 \bar\theta^a \bar\theta^b \bar\theta^c \,\e_{abc} \bar{\mathbb{D}}(y)\,.
\end{align} 
From the Gra\ss mann expansion of the two-point function \eqref{2-point-superfield} we extract to the following two-point functions
\begin{align}\label{2-p-insertions}
    \langle \mathbb{F}(t_1)\bar{\mathbb{F}}(t_2) \rangle  &= \frac{C_\Phi}{t_{12}} &
     \langle \mathbb{O}^a(t_1)\bar{\mathbb{O}}_b(t_2) \rangle  &= \frac{2 \,c_\Phi\,\delta^a_b}{t_{12}^{2}} \\
      \langle \mathbb{\Lambda}^a(t_1)\bar{\mathbb\Lambda}_b(t_2) \rangle & = \frac{8 \,C_\Phi\,\delta^a_b}{t_{12}^{3}}&
     \langle \mathbb{D}(t_1)\bar{\mathbb{D}}(t_2) \rangle  &= \frac{12\, C_\Phi}{t_{12}^{4}}
\end{align}
 
For the case of the displacement supermultiplet the normalization factor $C_{\Phi}$ has an important physical interpretation. As we showed in section \ref{sec:weakcoupling} most of the components in the displacement multiplet are obtained by the action of a broken symmetry generator. Such generators have a natural normalization in the bulk theory and therefore the Ward identities \eqref{D} fix the physical normalization of the displacement operator, making its two-point function an important piece of defect CFT data. There is concrete evidence that in the presence of a superconformal defect this coefficient is related to the one-point function of the stress tensor operator \cite{Lewkowycz:2013laa,Bianchi:2018zpb,Bianchi:2019sxz}. Moreover, for the case of the Wilson line this coefficient is particularly important as it computes the energy emitted by an accelerating heavy probe in a conformal field theory \cite{Correa:2012at,Fiol:2012sg}, often called Bremsstrahlung function. The relation with the two-point function of the displacement operator in the context of superconformal theories in three and four dimensions has allowed for the exact computation of this quantity in a variety of examples~\cite{Correa:2012at,Fiol:2012sg,Lewkowycz:2013laa,Forini:2012bb,Correa:2014aga,Bianchi:2014laa,Fiol:2015spa,Mitev:2015oty,defectABJM,Bianchi:2017svd,Bianchi:2018scb,Bianchi:2018zpb,Fiol:2019woe,Bianchi:2019dlw}. In the present context we have
\be\label{Bremss}
C_{\Phi}(\lambda)=2 B_{1/2}(\lambda)\,,
\ee
where $B_{1/2}(\lambda)$ is the Bremsstrahlung function associated to the 1/2 BPS Wilson line in ABJM theory (see, e.g., \cite{Bianchi:2014laa,defectABJM}). In this paper we focus on the four-point correlators \eqref{4-point-1-primary} and \eqref{4-point-2-primary}, where the disconnected part has been factored out and the functions $f(z)$ and $h(\chi)$ are independent of the chosen normalization.

For their use below, we now extract some specific components of the correlator \eqref{4-point-1}, by expanding both sides in Gra\ss mann variables 
\allowdisplaybreaks
\begin{align}\label{corrF}
&\!\!\!\!\! \braket{\mathbb{F}(t_1)\bar{\mathbb{F}}(t_2)\mathbb{F}(t_3)\bar{\mathbb{F}}(t_4)}=\frac{C_{\Phi}^2}{t_{12} t_{34}}f(z)\\\nonumber
&\!\!\!\!\! \braket{\mathbb{O}^{a_1}(t_1)\bar{\mathbb{O}}_{a_2}(t_2)\mathbb{O}^{a_3}(t_3)\bar{\mathbb{O}}_{a_4}(t_4)}=\frac{4 C_{\Phi}^2}{t_{12}^2 t_{34}^2}\,\Big[\delta^{a_1}_{a_2} \delta^{a_3}_{a_4}\, \big(\,f(z)-z f'(z)+z^2f''(z) \big)\\\label{corrO}
 &\qquad\qquad\qquad\qquad\qquad\qquad\qquad \qquad  -\delta^{a_1}_{a_4} \delta^{a_3}_{a_2}\,\big(z^2f'(z)+z^3 f''(z)\big)\,\Big]
 \\\nonumber
&\!\!\!\!\! \braket{\mathbb{D}(t_1)\bar{\mathbb{D}}(t_2)\mathbb{D}(t_3)\bar{\mathbb{D}}(t_4)}=\frac{(12 C_{\Phi})^2}{t_{12}^4 t_{34}^4}\, 
 \frac{1}{36}\Big[36 f(z)-36 (z^4+z) f'(z) +18 z^2 (-14 z^3+3 z^2+1) f''(z)\\\nonumber
&\qquad\qquad\qquad\qquad  -6 z^3 \left(55 z^3-39 z^2+3 z+1\right) f^{(3)}(z)-3 z^4 \left(46 z^3-63 z^2+18 z-1\right) f^{(4)}(z)\\\label{corrD}
& \qquad\qquad\qquad \qquad -3 (z-1)^2 z^5 (7 z-1) f^{(5)}(z) -(z-1)^3 z^6 f^{(6)}(z) 
 \,\Big]\,\\\nonumber
&\!\!\!\!\! \braket{\mathbb{D}(t_1)\bar{\mathbb{D}}(t_2)\mathbb{O}^{a_3}(t_3)\bar{\mathbb{O}}_{a_4}(t_4)}=\frac{24 C_{\Phi}^2}{t_{12}^4t_{34}^2}\delta^{a_3}_{a_4}\frac16\Big[\,(\!1\!-z)\, z^4 f^{(4)}\!(z)-(3 z+\!1\!) \,z^3 f^{(3)}\!(z)\\\label{corrMIXED}
 &\qquad\qquad\qquad \qquad\qquad\qquad\qquad\qquad+3 z^2\, f''(z)-6 z f'(z)+6 f(z)\,\Big]\,,
\end{align}
where for each correlator we factorized the (squared) two-point function contribution~\eqref{2-p-insertions}  arising from the double OPE. 
In the next section we will use analytic bootstrap techniques to evaluate, at leading and subleading order at strong coupling,  the function $f(z)$. 
In Section~\ref{sec:sigmamodel} we will confirm the superspace analysis of this section by evaluating directly the correlators at strong coupling, using AdS/CFT,  via Witten diagrams. 
We will namely consider the correlation functions of the string excitations corresponding to the various defect operators and obtain that they are given in terms of (derivatives of) a uniquely determined function $f(z)$, verifying explicitly the relations above.

\section{Bootstrapping the supercorrelator}\label{bootstrap}
\label{sec:bootstrap}

In this section, by imposing symmetries and consistency with the OPE expansion, we extract the leading strong coupling correction to the four-point function of the displacement supermultiplet. We then extract the anomalous dimensions and the OPE coefficients of the composite operators appearing in the intermediate channels.  

Let us start from the expressions of the four-point functions of the superprimary \eqref{4-point-1-primary} and \eqref{4-point-2-primary}, which for this specific case read
\begin{align}
 \braket{\mathbb{F}(t_1)\bar{\mathbb{F}}(t_2)\mathbb{F}(t_3)\bar{\mathbb{F}}(t_4)}&=\frac{C_{\Phi}^2}{t_{12} t_{34}} f(z)\\
  \braket{\mathbb{F}(t_1)\bar{\mathbb{F}}(t_2)\bar{\mathbb{F}}(t_3)\mathbb{F}(t_4)}&=\frac{C_{\Phi}^2}{t_{12} t_{34}} h(\chi)
\end{align}
and we define
\begin{align}\label{fhatdef}
 \hat{f}(\chi)=\frac{f(\frac{\chi}{\chi-1})}{\chi}
\end{align}
such that the crossing equation simply reads
\begin{align}
 \hat{f}(\chi)=\hat{f}(1-\chi)\,.
\end{align}

\subsection{Leading order}\label{leadingorder}
The leading order result at strong coupling can be easily obtained from Wick contractions, and reads
\begin{align}
 f^{(0)}(z)&=1-z & 
 h^{(0)}(\chi)&=1-\chi
\end{align}
or, alternatively $\hat{f}^{(0)}(\chi)=\frac{1}{\chi(1-\chi)}$. Unsurprisingly, the two correlators have the same functional form. Looking at the $s$-channel expansion of the two correlators \eqref{superblock} and \eqref{hexpansion} and using the well-known fact that in free theory the only exchanged operators are of the schematic form $[\mathbb{F}\bar{\mathbb{F}}]_n\sim\mathbb{F}\pa_t^n \bar{\mathbb{F}}$ with dimension $h_n=1+n$, we immediately realize that the mismatch factor $(-1)^{1+n}$ between \eqref{hexpansion} and \eqref{superblock} is compensated by an identical factor from the relation~\eqref{ctildec} for the OPE coefficients, which in the case of the operators $[\mathbb{F}\bar{\mathbb{F}}]_n$ reads
\begin{align}\label{c0andctilde0}
c^{(0)}_n=(-1)^{1+n} \tilde c^{(0)}_n
\end{align}
As a result, in free theory the two expansions \eqref{superblock} and \eqref{hexpansion} are identical up to the exchange $z\leftrightarrow \chi$. It is also easy to extract the explicit form of the OPE coefficient at leading order  
\begin{align}
 c^{(0)}_n=\frac{\sqrt{\pi } 2^{-2 n-3} \Gamma (n+4)}{(n+1)\Gamma \left(n+\frac{5}{2}\right)}
\end{align}
In a similar fashion one can read off the form of the coefficients $\mathsf{c}^{(0)}_n$ appearing in \eqref{tildehexpansion}. As expected, in the OPE of two identical fermions, only operators of the schematic form $[\mathbb{F}\pa^n\mathbb{F}]_n$ with \emph{odd} values of $n$ appear. We find
\begin{align}
 \mathsf{c}^{(0)}_n&=\frac{\sqrt{\pi } 2^{1-2 n} \Gamma (n+1)}{\Gamma \left(n+\frac{1}{2}\right)}  & &n \text{ odd}\\
 \mathsf{c}^{(0)}_n&=0  & &n \text{ even}
\end{align}
In the following we would like to consider a first-order perturbation of this result.

\subsection{Next-to-leading order}
We are interesting in finding the first-order strong coupling correction to the correlator. We expand the function $\tilde f(\chi)$ as 
\begin{align}\label{fhat}
 \hat{f}(\chi)&=\hat{f}^{(0)}(\chi)+\e\, \hat{f}^{(1)}(\chi) & h(\chi)&=h^{(0)}(\chi)+\e\, h^{(1)}(\chi)
\end{align}
where $\e$ is a small parameter, whose precise relation with the string tension cannot be predicted by symmetry considerations. Following~\cite{Carlo, Ferrero:2019luz}, we start with the following Ansatz for the first order correction to $\hat{f}$
\begin{align}\label{ansatz}
 \hat{f}^{(1)}(\chi)=r(\chi) \log(1-\chi)+r(1-\chi) \log \chi+q(\chi)
\end{align}
where $r(\chi)$ and $q(\chi)$ are rational functions and 
\begin{align}\label{qcross}
 q(\chi)=q(1-\chi)
\end{align}
In Appendix \ref{detailsbootstrap} we show explicitly that comparing the expansions \eqref{superblock} and \eqref{hexpansion} one can express the function $h^{(1)}(\chi)$ in~\eqref{fhat} in terms of $r(\chi)$ and $q(\chi)$. Here we only report the final result for the function $\hat{h}^{(1)}(\chi)=\frac{1-\chi}{\chi} h^{(1)}(\chi)$ introduced in \eqref{tildeh}
\begin{align}\label{ansatzh}
 \hat{h}^{(1)}(\chi)=-r\left(\tfrac{1}{1-\chi}\right)\log \chi + \left[r\left(\tfrac{\chi}{\chi-1}\right)+r\left(\tfrac{1}{1-\chi}\right)\right]\log (1-\chi)-q\left(\tfrac{\chi}{1-\chi}\right)
\end{align}
The final result for $h^{(1)}(\chi)$ is essentially what you would obtain by making the transformation $\chi\to \frac{\chi}{\chi-1}$ in $\hat{f}^{(1)}$ and neglecting the imaginary part of the logarithm. This is equivalent to the prescription, given in~\cite{Giombi} of putting an absolute value in the argument of the logarithm, but a precise justification of this fact can only be obtained by comparing the block expansions \eqref{superblock} and \eqref{hexpansion} as we do in Appendix~\ref{detailsbootstrap}.

Now we would like to bootstrap the rational functions $r$ and $q$ imposing crossing and consistency with the block expansions \eqref{tildehexpansion} and \eqref{superblock}. First, we rewrite the latter for $\hat{f}(\chi)$ in the $s$- and $t$-channels
\begin{align}
 \hat{f}(\chi)&=\frac{1}{\chi} + \sum_\Delta c_\Delta \hat{G}_{\Delta}(\chi)\\
 \hat{f}(\chi)&=\frac{1}{1-\chi} + \sum_\Delta c_\Delta \hat{G}_{\Delta}(1-\chi)
\end{align}
with $\hat{G}_\Delta(\chi)=\chi^{-1} G_\Delta(\frac{\chi}{\chi-1})=\chi^{\Delta-1} {}_2F_1(\Delta,\Delta+3;2\Delta+3;\chi)$. We consider the perturbation
\begin{align}
 \Delta_n&=1+n+\e\gamma^{(1)}_n+\mathcal{O}(\e^2)\\
 c_n&=c^{(0)}_n+\e c^{(1)}_n+\mathcal{O}(\e^2)
\end{align}
which allows to write the expansion for $f^{(1)}(\chi)$
\begin{align}
 \hat{f}^{(1)}(\chi)&= \sum_n \left(c^{(1)}_n \hat{G}_{1+n}(\chi)+ c^{(0)}_n \gamma^{(1)}_n \pa_\Delta \hat{G}_{\Delta}(\chi)|_{h=1+n}\right)
\end{align}
This expansion is regular for $\chi\to 0$ and crossing obviously guarantees that $\hat f^{(1)}(\chi)$ is regular also at $\chi\to 1$. Staring at the Ansatz \eqref{ansatz} one immediately concludes that $r(\chi)$ is regular at $\chi\to 1$. Since we assume that $r(\chi)$ and $q(\chi)$ are rational functions and we expect them to have poles for physical values of $\chi$, i.e. $\chi=0$ and $\chi=1$, the most general form for the function $r(\chi)$ is 
\begin{align}\label{rexpansion}
 r(\chi)=\sum_{m=-M_1}^{M_2} r_m \chi^m\,
\end{align}
where $M_2$ and $M_1$ take integer values and $M_2\geq -M_1$. Notice that, in general, $r(\chi)$ can be singular at $\chi\to 0$ as it multiplies $\log(1-\chi)$ (which is regular in this limit) and the divergence can be canceled by a pole in the function $q(\chi)$. The latter must respect the symmetry \eqref{qcross} and its most general form is
\begin{align}\label{qexpansion}
 q(\chi)=\sum_{l=-L_1}^{L_2} q_l \chi^l (1-\chi)^l
\end{align}
for integer values of $L_1$ and $L_2$ and $L_2\geq -L_1$.
Imposing the aforementioned cancellation of poles between $r(\chi)$ and $q(\chi)$ one finds several constraints which we derive in Appendix \ref{detailsbootstrap}. The final result is that all the coefficients $r_m$ in the expansion \eqref{rexpansion} are fixed in terms of the coefficients $q_l$ in \eqref{qexpansion}. Furthermore, we find
\begin{align}
 M_1&=L_1+1 & M_2&=2L_2+1 
\end{align}
Therefore, we are left with an infinite number of solution parametrized by the $L_1+L_2+1$ coefficients $q_l$. In this respect, one can construct a minimal solution by keeping a single term in the sum \eqref{qexpansion}. One criterium to choose which term to keep is the analysis of the large $n$ behaviour of $\gamma^{(1)}_n$.  In higher-dimensional examples of AdS/CFT the large twist behaviour of the anomalous dimension of the so-called ``double trace operators'' (of the schematic form $\mathcal{O}\square^n (\pa)^\ell \mathcal{O}$) is related to the relevance of the bulk interaction and it is bounded from above \cite{Heemskerk:2009pn, Fitzpatrick:2011dm}. Nevertheless, there are at least two obstacle for the direct application of this argument to the one-dimensional case \cite{Carlo}. First, there is no distinction between the spin and the twist of the double trace operators, which are characterized, in one dimension, by a single parameter $n$ labeling the number of derivatives ($\mathcal{O}\pa^n\mathcal{O}$). Secondly, in the higher-dimensional case the large $N$ analysis allows to lift the degeneracy between double trace operators and other operators with identical quantum numbers at the classical level. Here, as we will discuss in great detail in section \ref{cftdata}, this degeneracy is not lifted, preventing us from a clear identification of the operators appearing in the tree-level OPE with the double-trace operators. Therefore, analogously to the $\mathcal{N}=4$ SYM case \cite{Carlo}, we use the large-$n$ behaviour of the anomalous dimension simply as an organizing principle for the infinite set of solutions. 
The term with the mildest large $n$ behaviour  has $l=-1$. All the negative values of $l$ give a large $n$ behaviour of the order $\gamma_n \sim n^{-2l}$, while for positive values of $l$ the growth is much faster (already $l=0$ gives a $\gamma_n\sim n^5$). We conclude that the functions $r(\chi)$ and $q(\chi)$ yielding the mildest behaviour at large $n$ are
\begin{align}\label{solutionrandq}
 r(\chi)&=\frac{r_{-2}}{\chi^2}+ \frac{r_{-1}}{\chi}  & q(\chi)&=\frac{q_{-1}}{\chi(1-\chi)}
\end{align}
As we show in Appendix \ref{detailsbootstrap} all the $r_m$ coefficients can be fixed in terms of the $q_l$. In this specific case we get
\begin{align}\label{randqcoeff}
 r_{-2}&=q_{-1} & r_{-1}=2q_{-1} 
\end{align}
It is instructive, though, to see how this result can be obtained in this very simple situation. The first relation in \eqref{randqcoeff} simply arises by requiring that the pole at $\chi=0$ in the function $q(\chi)$ is canceled by the product $r(\chi)\log(1-\chi)$. To fix the coefficient $r_{-1}$, instead, one needs to analyse the function $\hat{h}(\chi)$ in \eqref{ansatzh} and impose consistency with the block expansion \eqref{hexpansion}. In the chiral-chiral channel we perturb the leading order result by
\begin{align}
 \Updelta_n&=1+n+\e \upgamma^{(1)}_n + \mathcal{O}(\e^2)& &n \text{ odd} \label{hnchiralchiral}\\
 \mathsf{c}_n&=\mathsf{c}^{(0)}_n+\e \mathsf{c}^{(1)}_n+ \mathcal{O}(\e^2) &  &n \text{ odd}
\end{align}
In \eqref{boundexchange}, we showed that unprotected exchanged operators in the chiral-chiral channel must have dimension $\D>3$. Therefore, in \eqref{hnchiralchiral} we must impose $\upgamma_1=0$. This is translated in the absence of terms like $(1-\chi)^2 \log(1-\chi)$ in $\hat{h}(\chi)$. Using the Ansatz \eqref{ansatzh} with the solution \eqref{solutionrandq} one immediately finds $r_{-1}-2 r_{-2}=0$. The presence of a protected operator of dimension $2$ of the form $\mathbb{F}\pa \mathbb{F}$ might be surprising and one may wonder what is the reason for the large gap \eqref{boundexchange} for the unprotected spectrum. Since the origin of this large gap is superconformal symmetry, one may answer this question by looking at the correlators of superdescendants. This analysis shows that not imposing the absence of the anomalous dimension $\upgamma_1$ translates in the appearance of unphysical exchanged operators in the superdescendant block decompositions. 

The remaining overall factor $q_{-1}$ can be reabsorbed in the definition of $\epsilon$ leaving us with the final solution for $\hat{f}^{(1)}(\chi)$
\begin{align}
 \hat{f}^{(1)}(\chi)=-\left(\frac{1}{\chi^2}+\frac{2}{\chi}\right) \log (1-\chi)- \left(\frac{1}{(1-\chi)^2}+\frac{2}{1-\chi}\right) \log (\chi)-\frac{1}{\chi(1-\chi)}
\end{align}
It is useful to write down also the explicit expressions of $f^{(1)}(z)$ and $\hat{h}^{(1)}(\chi)$, which can be simply obtained from \eqref{fhatdef} and \eqref{ansatzh} (see also Appendix \ref{detailsbootstrap})
\begin{align}
 f^{(1)}(z)&=-\frac{(1-z)^3}{z} \log (1-z)+z(3-z)\log(-z)+z-1 \label{finalresultfz}\\
 \hat{h}^{(1)}(\chi)&=\frac{\chi-1}{\chi} \left[\frac{(1-\chi)^3}{\chi} \log (1-\chi)-\chi(3-\chi)\log(\chi)+1-\chi\right] \label{finalresulthtilde}
\end{align}
where the overall sign has been chosen to match the string theory analysis of section \ref{sec:sigmamodel} with $\e>0$. Using this result, it is a straightforward exercise to extract the defect OPE data. We summarize our results in the next section.

\subsection{Extracting CFT data}\label{cftdata}

Before computing the values of anomalous dimensions and OPE coefficients, we need to comment on the class of operators one expects to appear in this context. When classifying operators that can be exchanged in a given channel, one is interested in eigenstates of the dilatation operators. When perturbing the leading order result, one can think of building operators out of the fundamental fields in the worldsheet lagrangian (see Section \ref{sec:sigmamodel}), which are in one-to-one correspondence with the components of the superdisplacement multiplet. In this respect, it is easy to realize that ``two-particle'' operators of the schematic form $O_n\sim\mathbb{F}\pa^n\bar{\mathbb{F}}$ will have a three-point function $\braket{\mathbb{F}\bar {\mathbb{F}} O_n}$ that is leading compared to higher particle operators. Nevertheless, these two-particle operators are not well defined eigenstates of the dilatation operator. A simple example is the mixing between two- and four-particle operators $\mathbb{F}\pa^2 \bar{\mathbb{F}}$ and $\mathbb{F} \bar{\mathbb{F}}\mathbb{F} \bar{\mathbb{F}}$. Only a linear combination of these two operators will be an eigenstate of the one-loop dilatation operator, but both of them are allowed to appear in the $\mathbb{F}\times \bar{\mathbb{F}}$ OPE \footnote{We would like to thank Pietro Ferrero, Shota Komatsu and Carlo Meneghelli for very useful discussions on this point.}. Of course the situation becomes increasingly more intricate for heavier operators. Therefore, we are led to conclude that any operator which includes a two-particle contribution will appear in the leading order OPE and the anomalous dimension we extract is actually a linear combination of the anomalous dimensions of these operators, weighted by their OPE coefficients. Given that part of this degeneracy is lifted by decomposing in terms of superconformal blocks instead of ordinary conformal blocks there are a few cases, where we can be certain that a single long multiplet with a given dimension can appear. These are the $n=0$ and $n=1$ case for the $\mathbb{F}\times \bar{\mathbb{F}}$ channel associated to the operators $\mathbb{F}\bar{\mathbb{F}}$ and $\mathbb{F}\pa\bar{\mathbb{F}}$ and the $n=1$ case in the $\mathbb{F} \times \mathbb{F}$ channel associated to $\mathbb{F}\pa\mathbb{F}$, which is protected. To solve the mixing for heavier operators one would have to study a larger class of correlators, a task which goes beyond the scope of this paper.

Given this caveat, we are ready to extract the values of anomalous dimensions and OPE coefficients in the two channels. Let us start from the chiral-antichiral channel. Comparing the expansion \eqref{oneloopexpansionf} with our result \eqref{finalresultfz} one finds
\begin{align}\label{andimexpansion}
 z(3-z)=\sum_{n\geq0} c_n^{(0)} \gamma^{(1)}_n G_{1+n}(z)
\end{align}
The $G_{1+n}$ functions, for integer $n$, form an orthonormal basis of solutions of the differential equation $\mathcal{L}G_\lambda=\lambda G_\lambda$ with $\mathcal{L}=-z^2(z-1)\partial_z^2 -z(z-3)\partial_z$ in \eqref{eigenfunction}. Therefore one can derive the orthogonality relation \footnote{The weighted inner product is obtained requiring the differential operator $\mathcal{L}$ to be self-adjoint~$\langle G_{\lambda_1}, \mathcal{L} G_{\lambda_2}\rangle_w=\langle \mathcal{L} G_{\lambda_1}, G_{\lambda_2}\rangle_w$, which is translated in the equation~$\mathcal{L}^{-1}w-\mathcal{L}w=0$ for the weight $w$,  solved by $w=\frac{z}{(1-z)^3}$ which also ensures orthonormality.}
\begin{equation}\label{orthogonality}
    \oint \frac{dz}{2 \pi i} \frac{z}{(1-z)^3} G_{1+n}(z) G_{-3-n'}(z)= \delta_{n,n'}\,,
\end{equation}
where the contour circles $0$  counterclockwise. Using this to invert the expansion \eqref{andimexpansion} we obtain
\begin{align}
 \gamma^{(1)}_n&=\frac{1}{c_n^{(0)}}\oint \frac{dz}{2\pi i}\, \frac{z^2 (3-z)}{(1-z)^3} G_{-3-n}(z)
\end{align}
Solving the integral we get
\begin{align}
 \gamma^{(1)}_n&=-n^2-4n-3
 \end{align}
The correction to the OPE coefficients can be easily obtained at each value of $n$ finding agreement with the general relation~\cite{Heemskerk:2009pn, Fitzpatrick:2011dm,Alday:2014tsa} 
 \be
  c_n^{(1)}=\frac{\partial}{\partial n} \left(c^{(0)}_n\,\gamma^{(1)}_n\right)
\ee
Their explicit expression is
 \begin{align}
 c_n^{(1)}&=c_n^{(0)} \left[-2-4n+ \gamma^{(1)}_n \left(\psi(n+4)- \psi(n+\tfrac{5}{2})-2\log (2)-\frac{1}{n+1} \right)\right]
 \end{align}
where $\psi(n)=\frac{\Gamma'(n)}{\Gamma(n)}$. 

A similar analysis can be carried out in the chiral-chiral channel. In this case the expansion is in terms of ordinary conformal blocks and we have the orthogonality relation 
\begin{align}
 -\oint \frac{d\chi}{2\pi i} \frac{1}{(1-\chi)^2} g_{1+n}(1-\chi) g_{-n'}(1-\chi)=\delta_{n,n'}
\end{align}
where the contour is a circle around $\chi=1$. We use it to invert the OPE expansion
\begin{align}
-\frac{(1-\chi)^4}{\chi^2}=\sum_{n\geq3} \mathsf{c}_n^{(0)} \upgamma^{(1)}_n g_{1+n}(1-\chi)
\end{align}
obtained comparing the coefficient of $\log(1-\chi)$ in \eqref{finalresulthtilde} with the perturbative expansion of \eqref{tildehexpansion}. This gives
\begin{align}
 \upgamma^{(1)}_n&=-n^2-n+2\,,  & &n \text{ odd}\,.
\end{align}
As expected, $\upgamma^{(1)}_1=0$. Also in this case, for the OPE coefficients we have
\begin{align}
 \mathsf{c}^{(1)}_n&= \pa_n (\mathsf{c}^{(0)}_n \upgamma^{(1)}_n)\,,
\end{align}
which gives
\begin{align}
 \mathsf{c}^{(1)}_n&=\mathsf{c}^{(0)}_n\left[ -1-2n +\upgamma^{(1)}_n \left(\psi(n+1)-\psi (n+\tfrac12)-2 \log 2\right) \right] & &n \text{ odd}\,.
\end{align}

\section{Correlators from Witten diagrams in AdS$_2$}
\label{sec:sigmamodel}

The Type IIA background $\ads_4 \times\cp^3$ is defined via
\begin{gather}
  \label{total_metric}
ds^2=R^2\left ( ds^2_{\textrm{AdS}_4}+ 4 ds^2_{\cp^3}\right) \,,
  \qquad
  e^\phi= \frac{2 R}{ k} \,,
  \qquad
  R^2 \equiv {\widetilde{R}^3\over 4 k}= \frac{k^2}{4} e^{2\phi} \,,
  \\
  F_2 = 2\,k\,J_{\cp^3} \,,
  \qquad
  F_4=\frac{3}{8} R^3\, \vol(\ads_4) \,,
\end{gather}
where $R$ is the $\ads_4$ radius, $\phi$ the dilaton, $k$  results from the compactification of the original M-theory on $\ads_4\times \sphere^7/Z_k$ background and coincides in the dual theory with the  Chern-Simons  level number~\cite{ABJM}. Above, $F_2$, $F_4$ are the 2-form and 4-form field strengths with  $J_{\cp^3}$ the K\"ahler form on $\cp^3$.  As they only play a role in the 
fermionic part of the Lagrangian and we limit our analysis to the bosonic part,  we report them here only for completeness.   
Using a Poincar\'e patch for the AdS$_4$ metric
\be 
ds^2_{\ads_4} = \frac{dz^2 + dx^r dx^r }{z^2} \,,
\ee
where $x^r=(x^0,x^1,x^2)$ parametrize the Euclidean three-dimensional  boundary of $AdS_4$ and $z$ is the radial coordinate,
the bosonic part of the superstring  action in AdS$_4 \times \,\mathbb{CP}^3$   reads 
\begin{equation}\label{2.1} 
S_B =\frac{1}{2} T   \int d^2\s \sqrt{h}\,   h^{\m\n} \Big[  \frac{1}{z^2}  \left(\partial_\m x^r\partial_\n x^r+\partial_\m z\partial_\n z\right)
+   4\,G_{MN}^{\cp^3}\,{\partial_\m Y^M\partial_\n Y^N } \Big]  \ .
\end{equation}
Here, $\sigma^\mu= (t,s) $ are  Euclidean world-sheet  coordinates and $T$ is the effective string tension. In its original ``dictionary'' 
proposal~\cite{ABJM} it is related to the effective 't Hooft 
coupling $\lambda$ of the dual $\mathcal{N}=6$ superconformal Chern-Simons theory (realized in the limit of $k$ and $N$ large with  their ratio fixed)  via
\be\label{T}
T= 
\frac{R^2}{2\pi\alpha'}
=\sqrt{\frac{\lambda}{2}}\,,
\qquad\qquad
\lambda=\frac{N}{k}\,.
\ee
In fact, as we are interested at leading, tree-level order in perturbation theory we may disregard the corrections  to the effective string tension $T$  due to the geometry of the background~\cite{ABJM,Bergman:2009zh}, which start at order $\frac{1}{\sqrt{\lambda}}$~\cite{Bianchi:2014ada}.
The classical solution to \eqref{2.1} which is relevant here is the minimal surface  corresponding to the straight Wilson line at the boundary  
\be \label{solution}   
z= s \ , \ \ \ \ \ \ \  x^0 = t \ , \ \ \ \ \   \ \ \    \qquad x^i=0\   ,\qquad  \,i=1,2\,,
\ee
with all the remaining ($\cp^3$) coordinates vanishing.  This is just the straightforward embedding in the AdS$_4$ background of the 
solution of~\cite{dgt,Giombi}. 
The induced metric is the AdS$_2$ metric
\be
g_{\mu\nu} d\sigma^\mu d\sigma^\nu=   {1\ov s^2} ( dt^2 + ds^2)\,.
\ee
 We will consider correlators of small fluctuations  of  ``transverse"  string coordinates (the $x^i, \, i=1,2$ and the $\cp^3$ coordinates) near this minimal surface.
The bosonic symmetry of the defect conformal field theory associated to the 1/2 BPS Wilson line is SU$(1,1) \times $ SU$(3) \times $U$(1)_{J_0}$, and it turns out to be the manifest symmetry of the bosonic string action \eqref{2.1}. 
The SU$(1,1) \simeq$ SO$(2,1)$  symmetry 
can be made manifest by fixing  a static gauge where $z$ and $x^0$  do not fluctuate and using the following parametrization for the   embedding of AdS$_2$ into AdS$_4$~\cite{Giombi} 
\be \label{2.3} 
ds^2_4 =\frac{ (1+\frac{1}{2} |X|^2)^2}{(1-\frac{1}{2} |X|^2)^2} ds^2_{2}  + \frac{dX d\bar X}{(1-\frac{1}{2} |X|^2)^2} \ , \ \ \ \ \ \  \ \ \ \ \ \
ds^2_2 =\frac{1}{z^2} (dx_0^2+dz^2)      \ . 
\ee
where we introduced the complex combination  $X=\frac{1}{\sqrt{2}}(x^1+i x^2)$ in terms of the transverse AdS coordinates $x^i$~. $X$ and $\bar X$ have opposite charge under  $U(1)_{J_0}$.   
Finally, adopting the following parametrization of the $\cp^3$ metric
\be
ds^2_{\cp^3}=\frac{d\bar{w}_a\,dw^a}{1+|w|^2}-\frac{d\bar{w}_a\,w^a\,d{w}^b\,\bar w_b}{(1+|w|^2)^2} \,, \qquad |w|^2=\bar{w}_a \,w^a\,,\qquad a,b=1,2,3\,,
\ee 
the preserved SU$(3)$ subgroup of the SU$(4)$ global symmetry of $\cp^3$ is manifest.  

The  Nambu-Goto action with fixed static gauge reads then
reads
%
\begin{equation}\label{stringaction}
   S_B = T \int d^2\sigma \sqrt{\det\Big[ \frac{(1+1/2|X|^2)^2}{(1-1/2|X|^2)^2} g_{\mu \nu} + 2 \frac{\del_\mu X \del_\nu \bar{X}}{(1-1/2|X|^2)^2} +4 \Big(\,\frac{\del_\mu \bar{w}_a \del_\nu w^a}{1+\abs{w}^2}-\frac{\del_\mu \bar{w}_aw^a \bar{w}_b \del_\nu w^b}{(1+\abs{w}^2)^2}\Big)\,\Big]} 
\end{equation}
 where $ g_{\m\n}= {1\ov s^2} \delta_{\m\n}$ is the  background AdS$_2$ metric. 
Along the lines of~\cite{Giombi}, \eqref{stringaction} can be interpreted as the action of a  straight fundamental   string in AdS$_4 \times \cp^3$ 
 stretched from the boundary towards the  AdS$_4$ center (so, stretched along  $z$), as well as the action for a 2d  ``bulk'' field  theory of  1+3 complex   scalars in AdS$_2$ geometry 
 with SO$(2,1) \times [$U$(1) \times  $SU$(3)]$ as manifest  symmetry. From the AdS/CFT point of view, this second interpretation leads to a 
CFT$_1$  dual  living at  the $z=s=0$ boundary, namely the  defect CFT defined by operator insertions on the straight Wilson line. 
  
Expanding the action above in powers of  $X$ and $w^a$ one gets
 \begin{align}   
S_B \equiv& T \int d^2\s   \sqrt{ g}\  L_B\,,\qquad L_B =\  L_2  + L_{4X}   + L_{2X,2w} + L_{4w}   + ...   \ , \la{lagr}\\
L_2=&\tet   g^{\m\n}\del_\m X \del_\n \bar{X}  +  2 |X|^2 + g^{\m\n}\del_\m w^a \del_\n \wb_a\ ,  \la{2.6} 
\\[2pt]
L_{4X} =&\ \tet  2|X|^4+|X|^2  \,(g^{\m\n}\del_\m X \del_\n \bar{X}) -\frac{1}{2}\,(g^{\m\n}\del_\m X \del_\n X) \,(g^{\r\k}\del_\r \Xb \del_\k \Xb) \,,
\label{2.7}            
\\[2pt]\nonumber
L_{2X,2w}=&\ \tet   (g^{\m\n}\del_\m X \del_\n \Xb  )\,(g^{\rho\kappa} \del_\rho w^a \del_\kappa  \wb_a) 
          -  (g^{\m\n} \del_\m X \del_\n w^a) \; (g^{\rho\kappa} \del_\rho \Xb \del_\kappa  \wb_a)\\
         &   -  (g^{\m\n} \del_\m \Xb \del_\n w^a) \; (g^{\rho\kappa} \del_\rho X \del_\kappa  \wb_a)\ ,  
\\[2pt]\nonumber
L_{4w} =&\ - \tet \frac{1}{2} \tet (w^a \wb_a) (g^{\m\n} \del_\m w^b \del_\n \wb_b) -  \frac{1}{2} \tet (w^a \wb_b) (g^{\m\n} \del_\m w^b \del_\n \wb_a) 
+   \frac{1}{2}\,(g^{\m\n}\del_\m w^a \del_\nu \wb_a)^2\\
&\tet-\frac{1}{2}  (g^{\m\n} \del_\m w^a \del_\n \wb_b) \; (g^{\rho\kappa} \del_\rho \wb_a \del_\kappa  w^b)
-\frac{1}{2}  (g^{\m\n} \del_\m w^a \del_\n w^b) \; (g^{\rho\kappa} \del_\rho \wb_a \del_\kappa  \wb_b)\ . \label{lagrend}
\end{align}
There are therefore  one  massive  ($X$ with $m^2=2$)   and  three  massless  ($w^a, ~a=1,2,3$) complex scalar fields propagating in AdS$_2$, that correspond to the bosonic elementary CFT$_1$ insertions represented in the displacement supermultiplet - respectively, to the $\Delta=2$ displacement operator $\mathbb{D}$ and to the $\Delta=1$ operators $\mathbb{O}^a,  ~a=1,2,3$.  In fact, as written above in Section~\ref{sec:superconformal}, to obtain the AdS/CFT dual of the full displacement supermultiplet one has to consider also the fermionic fluctuations. At quadratic level, the fermionic spectrum has been worked out in~\cite{Forini:2012bb, Aguilera-Damia}, and consists of two massless  and six massive fermions (of which three with mass $m_F=1$ and three with $m_F=-1$) which should correspond, respectively,   to the $\Delta=\frac{1}{2}$ fermionic superprimary $\mathbb{F}$ of the multiplet and its conjugate and to the $\Delta=\frac{3}{2}$ fermionic operators $\mathbb\Lambda_a$ and their conjugates.  Expanding the full Type IIA Green-Schwarz action in $\ads_4 \times\cp^3$ background~\cite{Gomis:2008jt,Uvarov:2009hf} around the solution~\eqref{solution} up to quartic order in fermions would yield  the interaction vertices from which to evaluate directly, via Witten diagrams, the four-point functions of fermionic fluctuations. Below we will limit our analysis to the direct calculation of bosonic four-point functions from the vertices in \eqref{lagr} above, and compare with the superspace results  of Section \ref{sec:superspace}. We emphasize however that in so doing we will in fact evaluate directly  the function $f(z)$ which governs the four-point correlator \eqref{corrF}  of the fermionic superprimary  $\mathbb{F}$ - and thus \emph{all} four-point functions - as the unique solution of the differential equations in \eqref{corrF}-\eqref{corrMIXED}, arising from the Gra\ss mann-expansion of the correlator for the four chiral fields in superspace.

Below, we will use these vertices of the  AdS$_2$ bulk theory to compute the corresponding tree-level Witten diagrams in AdS$_2$, with bulk-to-boundary propagators ending at points ${t_n}$ on the boundary.
As in the AdS$_5\times S^5$ case, no cubic terms appear in the bosonic Lagrangian above, so that  at this level of perturbation theory the correlation functions are only a sum of 4-point ``contact"  diagrams with four bulk-to-boundary propagators.

\subsection{Four-point function of massless fluctuations in $\cp^3$}

Here we compute the tree-level 4-point Witten diagram of the $\cp^3$ fluctuations $w,\wb$  appearing in the AdS$_2$ action in \eqref{lagr}-\eqref{lagrend}. 
 As discussed above, these are AdS/CFT dual to the scalar operator insertions $\mathbb{O}^a, \bar{\mathbb{O}}_a, ~a=1,2,3$ with protected dimension $\Delta=1$. 

Due to the $SO(2, 1)$ conformal invariance the 4-point function is expected to take the general form
 \begin{equation}\label{4-p-w}
 \langle w^{a_1}(t_1)\,\wb_{a_2}(t_2)\,w^{a_3}(t_3)\,\wb_{a_4}(t_4)\rangle  =  \frac{\big[C_{w}(\lambda)\big]^2}{t_{12}^2 t_{34}^2}
G^{a_1\,a_3}_{a_2\,a_4}(\chi) \ , 
\end{equation}
Here, $\chi$ is the conformally invariant cross-ratio defined in~\eqref{z-chi}, and  we used for the two point function
\begin{equation}\label{2-point-w}
\langle w^{a_1}(t_1)\wb_{a_2}(t_2)\rangle = \delta^{a_1}_{a_2}\frac{C_{w}(\lambda)}{t_{12}^2}\,.
\end{equation}
The function $G^{a_1\,a_3}_{a_2\,a_4}(\chi)$ in~\eqref{4-p-w} does not depend on the normalization of the $w^a$ fields, and thus on~$C_{w}(\lambda)$.  One can of course choose $C_w(\lambda)\equiv 4 B_{1/2}(\lambda)$, so to realize
a direct identification of $w^a$ with the $\mathbb{O}^a$ in view of~\eqref{2-p-insertions}~\footnote{This is the formal choice of~\cite{Giombi}, where the analogue relation is to the $\mathcal{N}=4$ SYM Bremsstrahlung function~\cite{Correa:2012at,Forini:2010ek,Drukker:2011za}. See also a related discussion in~\cite{Drukker:2020swu}.}. This would just correspond to an overall rescaling of the fields~\footnote{Given  the leading strong coupling value of the Bremsstrahlung function 
$B_{1/2}(\lambda)=\frac{\sqrt{2\lambda}}{4\pi}\equiv \frac{T}{2\pi}$, and given our choice~\eqref{bulk-to-boundary} of the bulk-to-boundary propagator, at tree level this would  amount to the rescaling $w^a\rightarrow \sqrt{2T} w^a$.}. 
By evaluating perturbatively the two-point function~\eqref{2-point-w} (namely, calculating loop corrections to the boundary-to-boundary propagator) one should then be able to verify that the elementary excitations $w^a$ are protected, as well as reproduce the strong coupling expansion of the corresponding 1/2 BPS Bremsstrahlung function~\eqref{Bremss}.

\begin{figure}
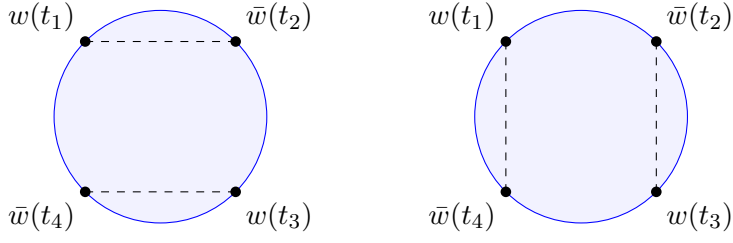

	\centering
\begin{wittendiagram}
 \draw[dashed] (-1,1) node[vertex] -- (1,1) node[vertex];
  \draw[dashed] (-1,-1) node[vertex] -- (1,-1) node[vertex];
    \node[anchor=south east] at (-1,1) {$w(t_1)$};
  \node[anchor=south west] at (1,1) {$\bar{w}(t_2)$};
  \node[anchor=north west] at (1,-1) {$w(t_3)$};
  \node[anchor=north east] at (-1,-1) {$\bar{w}(t_4)$};
\end{wittendiagram} \quad \quad \quad \begin{wittendiagram}
 \draw[dashed] (-1,1) node[vertex] -- (-1,-1) node[vertex];
  \draw[dashed] (1,1) node[vertex] -- (1,-1) node[vertex];
    \node[anchor=south east] at (-1,1) {$w(t_1)$};
  \node[anchor=south west] at (1,1) {$\bar{w}(t_2)$};
  \node[anchor=north west] at (1,-1) {$w(t_3)$};
  \node[anchor=north east] at (-1,-1) {$\bar{w}(t_4)$};
\end{wittendiagram}
\caption{Witten diagram for the disconnected contribution to the four-point function~\eqref{4-p-w}.}
   \label{fig:Witten-disconn}
\end{figure}
\begin{figure}
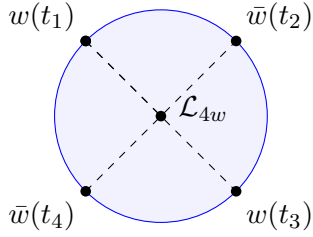

    \centering
\begin{wittendiagram}
 \draw[dashed] (-1,1) node[vertex] -- (1,-1) node[vertex];
  \draw[dashed] (-1,1) node[vertex] -- (0,0) node[vertex];
 \draw[dashed] (-1,-1) node[vertex] -- (1,1) node[vertex];
     \node[anchor=south east] at (-1,1) {$w(t_1)$};
  \node[anchor=south west] at (1,1) {$\bar{w}(t_2)$};
  \node[anchor=north west] at (1,-1) {$w(t_3)$};
  \node[anchor=north east] at (-1,-1) {$\bar{w}(t_4)$};
  \node[anchor=west] at (0.1,0.1) {$\mathcal{L}_{4w}$};
\end{wittendiagram}
    \caption{Witten diagram for the connected contribution to the four-point function~\eqref{4-p-w}.}
    \label{fig:Witten-conn}
\end{figure}

The disconnected part of the four-point function~\eqref{4-p-w} originates from Wick contractions, see  Fig~\ref{fig:Witten-disconn}, and reads
\begin{align}\nonumber
    \cor{w^{a_1}(t_1)\,\wb_{a_2}(t_2)\,w^{a_3}(t_3)\,\wb_{a_4}(t_4)}_\text{disconn.} &= \big[C_w(\lambda)\big]^2\,\Big[\frac{\delta^{a_1}_{a_2}\delta^{a_3}_{a_4}}{t_{12}^2 t_{34}^2}+\frac{\delta^{a_1}_{a_4}\delta^{a_3}_{a_2}}{t_{14}^2 t_{23}^2}\Big]  
    \\\label{w-disconn}
    &  =\frac{\big[C_{w}(\lambda)\big]^2}{t_{12}^2 t_{34}^2}\,\Big[\delta^{a_1}_{a_2}\delta^{a_3}_{a_4}+\frac{\chi^2}{(1-\chi)^2} \,\delta^{a_1}_{a_4}\delta^{a_3}_{a_2}\,\Big]\,.
    \end{align}
%
%
%
%
%
%
%
%
%
%
%
The first connected contribution to the four-point function  comes from the   tree-level connected Witten diagrams obtained from the four-point interaction vertices  $L_{4w}$ in~\eqref{lagrend}  with four bulk-to-boundary propagators attached, see Fig.~\ref{fig:Witten-conn}, and therefore it is subleading in $1/T$.  
As in~\cite{Giombi}, we will adopt the normalization~\cite{Penedones:2010ue, Fitzpatrick:2011ia} of the bulk-to-boundary propagator in dimension $d=1$   
\begin{equation}\label{bulk-to-boundary}
\!\!
K_{\Delta}(z,t;t') = {\cal C}_{\Delta} \Big[\frac{z}{z^2+(t-t')^2}\Big]^\Delta \equiv  {\cal C}_{\Delta}\,  \tilde{K}_{\Delta}(z,t;t')\ , \,\qquad {\cal C}_{\Delta} =\frac{\Gamma\left(\Delta\right)}{2\,\sqrt{\pi}\,\Gamma\left(\Delta+{1\ov 2}\right)} \,,
\end{equation}
for which the tree-level two-point function of the dual boundary operator is~$\langle O_{\Delta}(t_1)O_{\Delta}(t_2) \rangle ={ {\cal C}_{\Delta}\ov  t_{12}^{2\Delta}}$. 
Then the connected correlator reads 
\begin{equation}
\langle w^{a_1}(t_1)\,\wb_{a_2}(t_2)\,w^{a_3}(t_3)\,\wb_{a_4}(t_4)\rangle_\text{conn} = \frac{1}{T}\,\big(\mathcal{C}_{\Delta=1}\big)^4 \, \Big[\mathcal{Q}_1\, \delta^{a_1}_{a_2} \delta^{a_3}_{a_4}+\mathcal{Q}_2\,\delta^{a_1}_{a_4} \delta^{a_3}_{a_2}\Big]
\end{equation}
where ${\cal C}_{\Delta=1}={1\ov \pi}$ and $\mathcal{Q}^{a_1\,a_2}_{a_3\,a_4}$ is built out of the  $D$-functions~\cite{Liu:1998ty,DHoker:1999kzh, Dolan:2003hv} reviewed in Appendix~\ref{app:D-functions}, explicitly
\begin{eqnarray}\nonumber
\mathcal{Q}_1&=& \,3D_{1111}+t_{12}^2 D_{2211}+t_{34}^2 D_{1122}-2t_{24}^2 D_{1212}-2t_{13}^2 D_{2121}\\
&&  -3 t_{23}^2 \,D_{1221} - 3 t_{14}^2 D_{2 1 1 2}+4 (t_{14}^2 t_{23}^2 + t_{13}^2 t_{24}^2 - t_{12}^2 t_{34}^2) D_{2222}\,,\\\nonumber
\mathcal{Q}_2&=&\,3D_{1111}+t_{14}^2 D_{21 1 2} + t_{23}^2 D_{1 2 2 1} -  2 t_{13}^2 D_{2 1 2 1} -2 t_{24}^2 D_{12 1 2}\\
&&   -  3 t_{12}^2 D_{2 2 1 1} - 3 t_{34}^2  D_{1 1 2 2} +
 4 (  t_{12}^2 t_{34}^2+t_{13}^2 t_{24}^2-t_{14}^2 t_{23}^2 ) D_{2 2 2 2}\,.
\end{eqnarray}
This can be written in a conformally invariant way using the reduced $D$-functions $\bar D$ defined in \eqref{Dbar} - for which we also list explicit expressions in~\eqref{Dbar-explicit}-\eqref{Dbar-explicit-end}, leading eventually to the expression
\be\label{w-conn}
\!\!\!\!\!\!    \cor{w^{a_1}(t_1)\,\wb_{a_2}(t_2)\,w^{a_3}(t_3)\,\wb_{a_4}(t_4)}_\text{conn.} &=&\epsilon\,\frac{\big[\mathcal{C}_{\Delta=1}\big]^2}{ t_{12}^2 t_{34}^2}\,\Big[\delta^{a_1}_{a_2} \delta^{a_3}_{a_4}\,G_1(\chi)
    +\delta^{a_1}_{a_4} \delta^{a_3}_{a_2}\,G_2(\chi)\Big]\,
 \ee
 where, according to~\eqref{4-p-w},  we factored out a $[\mathcal{C}_{\Delta=1}\big]^2$ and the expansion parameter $\epsilon={\frac{1}{4\pi \,T}}$
 has been chosen to make contact with the bootstrap calculation. 
 
 For the functions $G_1$ and $G_2$ above, one finds
 \begin{eqnarray} \label{G1}
   G_1(\chi) &=&-3+\frac{1}{ (\chi-1 )}-\frac{\chi ^2 }{(1-\chi )^2}\log \chi +\left(1-\frac{4}{\chi   }\right) \log (1-\chi )\\\label{G2}
   G_2(\chi) &=&-\frac{\chi  (3 \chi +1)}{ (1-\chi )^2}+\frac{\chi ^2 (\chi +3)}{(\chi -1)^3}\,\log\chi-\log (1-\chi )
 \end{eqnarray}
Using~\eqref{z-chi-rel}, the functions above may be conveniently expressed in terms of the invariant $z$. Confronting then~\eqref{w-disconn} and~\eqref{w-conn} with~\eqref{corrO}  leads to the following system of second-order differential equations
\bal\label{system-f}
 f(z)-zf'(z)+z^2f''(z) &= 1+ \,\epsilon\big[-4+z-z^2\,\log (-z)+\big(z^2-\frac{4}{z}+3\big)\,\log(1-z)\,\big]\\
 -z^2f'(z)-z^3 f''(z) &=  z^2\,+\epsilon\,\big[\,z-4\,z^2-  z^2 (3-4z)\,\log (-z)+(1 + 3 z^2 - 4 z^3) \,\log (1-z)\,\big]\,
\eal
whose solution reads
\be\label{fstring}
f(z)=1-z+\epsilon\,\big[z-1+z (3-z) \log (-z)-\frac{(1-z)^3 }{z}\log (1-z)\big]\,,
\ee
in agreement with~\eqref{finalresultfz}.

One can now  repeat the analysis for the correlator $\cor{w^{a_1}(t_1)\,\wb_{a_2}(t_2)\,\wb^{a_3}(t_3)\,w_{a_4}(t_4)}$. The result coincides with the one obtained using on  the $f(z)$ above with the replacement $z\to \chi$, and neglecting the imaginary part of the logarithm.  This is in perfect agreement with what observed in the bootstrap analysis of Section~\ref{sec:bootstrap}~\footnote{See discussion below equation~\eqref{ansatzh}.} and discussed in Appendix~\ref{detailsbootstrap}. 
Below we will check that the $f(z)$ evaluated above solves also the corresponding differential equations  for the correlators of massive and mixed worldsheet excitations, once the normalization factors defining the corresponding two-point functions are identified with the ones of their field theory dual. 
 
\subsection{Four-point function of fluctuations in AdS$_4$}

The  complex field  $X$ appearing in the AdS$_2$  action in \eqref{lagr}-\eqref{lagrend} is the AdS/CFT dual  of the displacement operator insertion  $\mathbb{D}$, which has  protected dimension $\Delta=2$.    
Due to conformal invariance, the 4-point correlator reads then
 \begin{equation}
 \langle X(t_1)\,\Xb(t_2)\,X(t_3)\,\Xb(t_4)\rangle
  =  \frac{\big[C_{X}
(\lambda)\big]^2}{t_{12}^4 t_{34}^4}\,G(\chi) \ , 
\label{4point-X}
\end{equation}
where  we have used
\begin{equation}\la{4.2}
 \langle X(t_1)\,\Xb(t_2)\rangle=\frac{C_{{X}}(\lambda)}{t_{12}^4}\,.
\end{equation}
Again, the normalization factor $C_X(\lambda)$ may be chosen to be in direct correspondence with the Bremmstrahlung function~\eqref{Bremss} to realize a direct identification of $X$ with the displacement operator $\mathbb{D}$.   In view of~\eqref{2-p-insertions}, this would mean $C_X(\lambda)\equiv 24 B_{1/2}(\lambda)$.  

The   disconnected contribution to~\eqref{4point-X} reads 
\begin{equation}\label{X-disconn}
    \cor{X(t_1)\,\Xb (t_2)\,X(t_3)\,\Xb(t_4)}_\text{disconn.} =\frac{\big[C_{{X}}(\lambda)\big]^2}{t_{12}^4 t_{34}^4}\,\Big[1+\frac{\chi^4}{(1-\chi)^4}\,\Big]\,.
    \end{equation}
The subleading, connected contribution  is obtained evaluating Witten diagrams from the four-point interaction vertices $L_{4X}$ in \eqref{lagr}-\eqref{lagrend} and reads
\begin{eqnarray}
&&\langle X(t_1)\,\Xb (t_2)\,X(t_3)\,\Xb(t_4)\rangle_\text{conn} =\frac{1}{T}\,\big(\mathcal{C}_{\Delta=2}\big)^4 \, \mathcal{Q}\,, \qquad \mathcal{C}_{\Delta=2}=\frac{3}{2\pi}\\\nonumber
&&\qquad\mathcal{Q}=8 (D_{2 2 2 2} + t_{12}^2 D_{33 2 2}+ t_{34}^2 D_{2 2 3 3} +  t_{23}^2 D_{2 3 3 2} + t_{14}^2 D_{3 2 2 3}\\
&& \qquad\qquad~~ - 8 t_{24}^2 D_{23 2 3}- 
   8 t_{13}^2 D_{3 2 3 2}  +    16 t_{13}^2 t_{24}^2 D_{3 3 3 3})  \,.
\end{eqnarray}
Explicitly, and adopting the standard normalization, one gets
\begin{eqnarray}\nonumber
 \!\!\!   \cor{X\Xb X\Xb}_\text{conn.} \! \! \! &=&\epsilon\,\frac{\big[\mathcal{C}_{\Delta=2}\big]^2}{ t_{12}^4 t_{34}^4}\,
 \Big[\frac{115 \chi ^5-345 \chi ^4+543 \chi ^3-511 \chi ^2+246 \chi -48}{3 (1-\chi )^5}
 \\\label{X-conn}
   &&\qquad\qquad\qquad + \frac{2 \chi ^4 (5 \chi +3)}{(\chi -1)^5}\log\chi+2\big(\textstyle{5-\frac{8}{\chi }}\big) \log (1-\chi )
    \Big]\,.
 \end{eqnarray}
Writing everything in terms of  the invariant $z$, and confronting the superspace prediction \eqref{corrD} with \eqref{X-disconn} and \eqref{X-conn} we obtain 
\begin{eqnarray}\nonumber
&&\!\!\!\!\!\!\!
\frac{1}{36}\Big[36 f(z)-36 (z^4+z) f'(z) +18 z^2 (-14 z^3+3 z^2+1) f''(z) -6 z^3 \left(55 z^3-39 z^2+3 z+1\right) f^{(3)}(z)\\\nonumber
&&~~. -3 z^4 \left(46 z^3-63 z^2+18 z-1\right) f^{(4)}(z) -3 (z-1)^2 z^5 (7 z-1) f^{(5)}(z) -(z-1)^3 z^6 f^{(6)}(z) \,\Big]\\\nonumber
 &&\!\!\!\!=1+z^4+\epsilon\,\Big[\textstyle-16 -2 z-16 z^4-2 z^3-\frac{7 z^2}{3}
+2 (8 z-3) z^4 \log (-z)\\
&&\qquad\qquad\qquad\qquad \textstyle+\big( 6-\frac{16}{z}+6 z^4-16 z^5\big) \log (1-z)\Big]\,.
 \end{eqnarray}
This non-trivial sixth-order differential equation is immediately solved by the function $f(z)$ in~\eqref{fstring}.

Again, one may repeat the analysis for the correlator $\cor{X(t_1)\,\bar X(t_2)\,\bar X(t_3)\,X(t_4)}$. The result coincides with~\eqref{X-conn}
after the transformation $\chi\to \chi/(\chi-1)$, and neglecting the imaginary part of the logarithm.

\subsection{Four-point function of mixed fluctuations}

The 4-point correlator mixing two AdS $X$ fluctuations and two $\cp^3$ $w$ fluctuations reads
 \begin{equation}
\langle X(t_1)\,\Xb (t_2)\,w^{a_3}(t_3)\,\wb_{a_4}(t_4)\rangle
  =  \frac{C_{{X}}(\lambda)\,C_{{w}}(\lambda)}{t_{12}^4 t_{34}^2}\,\delta^{a_3}_{a_4}\,G(\chi) \ , 
\label{4point-mixed}
\end{equation}
The disconnected contribution is
\begin{equation}\label{mixed-disconn}
    \cor{X(t_1)\,\Xb (t_2)\,w^{a_3}(t_3)\,\wb_{a_4}(t_4)}_\text{disconn.} =\frac{C_{X}(\lambda) C_{w}(\lambda) }{t_{12}^4 t_{34}^2}\,\delta^{a_3}_{a_4}\,.
    \end{equation}
The connected contribution  is obtained evaluating Witten diagrams from the four-point interaction vertices $L_{4X}$ in \eqref{lagr}-\eqref{lagrend}, and reads
\begin{eqnarray}
&&\langle X(t_1)\,\Xb (t_2)\,w^{a_3}(t_3)\,\wb_{a_4}(t_4)\rangle_\text{conn} =\frac{1}{T}\,\big(\mathcal{C}_{\Delta=2}\big)^2 \big(\mathcal{C}_{\Delta=1}\big)^2 \, \delta^{a_3}_{a_4}\,\mathcal{Q}_{2X\,2w} \,, \\\nonumber
&&\qquad\mathcal{Q}_{2X\,2w}=4 \Big[\,D_{22 1 1} + 2 t_{12}^2 D_{33 1 1} + 2 t_{34}^2 D_{2 2 2 2} -   2 t_{24}^2 D_{2 3 1 2} - 2 t_{23} D_{2 3 2 1} \\
&& \qquad\qquad\qquad  \qquad - 2 t_{14}^2 D_{3 2 1 2} - 2 t_{13}^2 D_{3 2 2 1} +    4 (t_{14}^2 t_{23}^2 + t_{13}^2 t_{24}^2 - t_{12}^2 t_{34}^2) D_{3 3 2 2}\,\Big]\,,
\end{eqnarray}
 explicitly 
\begin{eqnarray} 
 \!\!\!   \cor{X(t_1)\,\Xb (t_2)\,w^{a_3}(t_3)\,\wb_{a_4}(t_4)}_\text{conn.} \! \! \! &=&\epsilon\,\frac{\mathcal{C}_{\Delta=2}\,\mathcal{C}_{\Delta=1}\, }{ t_{12}^4 t_{34}^2}\,\delta^{a_3}_{a_4}\,\Big[\frac{4 (\chi -2) \log (1-\chi )}{\chi }-8\Big]\,.
 \end{eqnarray}
Once again, writing everything in terms of the invariant $z$ and equating to the superspace prediction, the differential equation obtained
\be\nonumber
(\!1\!-z)\, z^4 f^{(4)}\!(z)-(\!1\!
-3 z) \,z^3 f^{(3)}\!(z)+3 z^2\, f''(z)+6 z f'(z)+6 f(z)
=1+\epsilon \big[-8+\textstyle{\frac{4 (z-2) }{z}}\,\log (1-z)\big]\\
\ee
is  solved by the function $f(z)$ in~\eqref{fstring}.

\section*{Acknowledgements}

We thank Fernando Alday, Lorenzo Di Pietro, Pietro Ferrero, Luigi Guerini, Shota Komatsu, Madalena Lemos, Marco Meineri, Carlo Meneghelli, Silvia Penati, Giulia Peveri and Edoardo Vescovi for useful discussions. 
The research of LB received funding from the European Union's Horizon 2020 research and innovation programme under the Marie Sklodowska-Curie grant agreement No~749909. The research of GB and VF has received funding from the European Union's Horizon 2020 research and innovation programme under the Marie Sklodowska-Curie grant agreement  No 813942.
The research of VF received funding from the STFC grant ST/S005803/1, from the Einstein Foundation Berlin through an Einstein Junior Fellowship, and was supported in part by Perimeter Institute for Theoretical Physics and the Simons Foundation through a Simons Emmy Noether Fellowship. 


\appendix
 \section{$osp(6|4)$ algebra and its subalgebra $su(1,1|3)$}\label{algebra} 
 \label{app:algebra-superconf}
We now list the commutation relations for the $osp(6|4)$ superalgebra. Let us start from the three-dimensional conformal algebra
\begin{align}
[P^{\mu},K^{\nu}]&=-2\d^{\mu\nu} D-2 M^{\mu\nu} & [D,P^{\mu}]&= P^{\mu} & [D,K^{\mu}]&=-K^{\mu}\\
 [M^{\mu\nu}, M^{\rho\sigma}]&=\d^{\sigma[\mu} M^{\nu]\rho}+\d^{\rho[\nu} M^{\mu]\sigma} & [P^{\mu},M^{\nu\rho}]&=\delta^{\mu[\nu}P^{\rho]} & [K^{\mu},M^{\nu\rho}]&=\delta^{\mu[\nu}K^{\rho]} 
\end{align}
Then we have the SU$(4)$ generators
\begin{align}
 [{J_I}^J,{J_K}^L]=\d_I^L {J_K}^J-\d^J_{K} {J_I}^L
\end{align}

Fermionic generators $Q^{IJ}_\a$ and $S^{IJ}_\a$ respect the reality condition $\bar Q_{IJ\a}=\frac12 \e_{IJKL} Q^{KL}_{\a}$ and similarly for $S$. Anticommutation relations are
\begin{align}\label{Qosp}
 \{Q^{IJ}_\a,Q^{KL\b}\}&=2i\,\e^{IJKL} {(\gamma^\mu)_{\a}}^{\b} P_\mu \qquad  \{S^{IJ}_\a,S^{KL\b}\}=2i\,\e^{IJKL} {(\gamma^\mu)_{\a}}^{\b} K_\mu \\  \{Q^{IJ}_\a,S^{KL\b}\}&=\e^{IJKL} ({(\gamma^{\mu\nu})_{\a}}^{\b} M_{\mu\nu}+2\d_{\a}^{\b} D)+2\d_{\a}^{\b}\e^{KLMN}(\d_M^J {J_N}^I-\d_M^I {J_{N}}^J)
\end{align}
Finally, mixed commutators are
\begin{align}
 [D,Q^{IJ}_{\a}]&=\frac12 Q^{IJ}_{\a} & [D,S^{IJ}_{\a}]&=-\frac12 S^{IJ}_{\a} \\
 [M^{\mu\nu},Q^{IJ}_{\a}]&=-\frac12 {(\gamma^{\mu\nu})_{\a}}^{\b} Q^{IJ}_{\b} & [M^{\mu\nu},S^{IJ}_{\a}]&=-\frac12 {(\gamma^{\mu\nu})_{\a}}^{\b} S^{IJ}_{\b}  \\
 [K^{\mu},Q_{\a}^{IJ}]&=-i\,{(\g^{\mu})_{\a}}^{\b} S^{IJ}_{\b} & [P^{\mu},S^{IJ}_{\a}]&=-i\,{(\g^{\mu})_{\a}}^{\b} Q^{IJ}_{\b}\\
 [{J_I}^J,Q^{KL}_{\a}]&=\d_I^K Q^{JL}_\a+\d_{I}^L Q^{KJ}_\a-\frac12\d_I^J Q^{KL}_{\a} &  [{J_I}^J,S^{KL}_{\a}]&=\d_I^K S^{JL}_\a+\d_{I}^L S^{KJ}_\a-\frac12\d_I^J S^{KL}_{\a}
\end{align}


Inside the $osp(6|4)$ it is possible to identify the $su(2|3)$ (or, more precisely $su(1,1|3)$) subalgebra preserved by the 1/2 BPS Wilson line. The $su(1,1)$ generators are those of the one-dimensional conformal group, i.e. $\{D,P\equiv P_1,K\equiv K_1\}$,
satisfying
\begin{align}
 [P,K]&=-2 D & [D,P]&=P & [D,K]&=-K
\end{align}
The SU$(3)$ generators ${R_a}^b$ are traceless, i.e. ${R_a}^a=0$ and they are given in terms of the original $su(4)$ ones by
\begin{align}
{R_{a}}^{b}&=\begin{pmatrix}
                             {J_2}^2+\frac13 {J_1}^1 & {J_2}^3 & {J_2}^4\\
                             {J_3}^2 & {J_3}^3+\frac13 {J_1}^1 & {J_3}^4\\
                              {J_4}^2 & {J_4}^3  & -{J_3}^3-{J_2}^2-\frac23 {J_1}^1\\
                            \end{pmatrix}             
\end{align}
Their commutation relations are
\begin{align}
 [{R_{a}}^{b},{R_{c}}^{d}]&=\d_{a}^{d} {R_{c}}^{b} -\d^{b}_{c} {R_{a}}^{d}
\end{align}
The last bosonic symmetry is the $u(1)$ algebra generated by
\begin{align}
J_0=3i M_{23}-2 {J_1}^1
\end{align}
and commuting with the other bosonic generators.

The fermionic generators are given by a reorganization of the preserved supercharges\newline $\{Q^{12}_+,Q^{13}_+,Q^{14}_+,Q^{23}_-,Q^{24}_-,Q^{34}_-\}$, together with the corresponding superconformal charges. Our notation is
\begin{equation}
 Q^a=Q^{1a}_{+} \qquad S^a=i\, S^{1a}_{+}  \qquad  \bar Q_a=i\,\frac12 \epsilon_{abc} Q_{-}^{bc} \qquad  \bar S_a=\frac12 \epsilon_{abc} S_{-}^{bc}
\end{equation}
The $i$ factors are chosen to compensate those in the algebra \eqref{Qosp}
so that 
\begin{align}
 \{Q^a,\bar Q_b\}&=2 \d^a_b P &  \{S^a,\bar S_b\}&=2\d^a_b K\\  \{Q^a,\bar S_b\}&= 2\d_b^a ( D+\tfrac13 J_0)-2 {R_b}^a & \{\bar Q_a, S^b\}&= 2\d_b^a(D-\tfrac13 J_0)+2 {R_a}^b 
\end{align}
Finally, non-vanishing mixed commutators are\small
\begin{align}
 [D,Q^a]&=\frac12 Q^a & [D,\bar Q_a]&=\frac12 \bar Q_a &  [K,Q^a]&=S^a & [K,\bar Q_a]&= \bar S_a\\
 [D,S^a]&=-\frac12 S^a & [D,\bar S_a]&=-\frac12 \bar S_a &  [P,S^a]&=-Q^a & [P,\bar S_a]&=- \bar Q_a\\
 [{R_a}^b,Q^c]&=\d_a^c Q^b-\tfrac13 \d_a^b Q^c & [{R_a}^b,\bar Q_c]&=-\d_c^b \bar Q_a+\tfrac13 \d_a^b \bar Q_c & [J_0,Q^a]&=\tfrac12 Q^a & [J_0,\bar Q^a]&=-\tfrac12 \bar Q^a\\
 [{R_a}^b,S^c]&=\d_a^c S^b-\tfrac13 \d_a^b S^c & [{R_a}^b,\bar S_c]&=-\d_c^b \bar S_a+\tfrac13 \d_a^b \bar S_c & [J_0,S^a]&=\tfrac12 S^a & [J_0,\bar S^a]&=-\tfrac12 \bar S^a
\end{align}\normalsize

\subsection{Representations of $su(1,1|3)$}\label{representations}
\label{app:reprs}
Here we present a summary of the representation theory of the $su(1,1|3)$ algebra. A detailed analysis can be found in \cite{defectABJM}. The algebra is characterized by four Dynkin labels $[\D,j_0,j_1,j_2]$ associated to the Cartan generators of the bosonic subalgebra $su(1,1)\oplus u(1)\oplus su(3)$. The two SU$(3)$ Cartan generators are defined as
\begin{equation}
 J_1={R_1}^1-{R_2}^2 \qquad J_2={R_1}^1+2{R_2}^2
\end{equation}
A highest weight state is characterized by
\begin{align}
 S^a\ket{\D,j_0,j_1,j_2}^{\text{hw}}&=0 & \bar S_a\ket{\D,j_0,j_1,j_2}^{\text{hw}}&=0 & E^+_a\ket{\D,j_0,j_1,j_2}^{\text{hw}}&=0
\end{align}
where $E^+_a$ are raising generators of SU$(3)$ in the Weyl Cartan basis (see \cite{defectABJM}). The long multiplet is built by acting with supercharges, momentum and SU$(3)$ lowering generators on the highest weight state. The dimension of the long multiplet is 
\begin{equation}
 \text{dim} \mathcal{A}^{\D}_{j_0;j_1,j_2}=27 (j_1+1)(j_2+1)(j_1+j_2+2)
\end{equation}
and unitarity requires
\begin{equation}\label{unitaritybounds}
 \D\geq\left\{\begin{array}{l}
           \frac13(2j_1+j_2-j_0) \quad j_0\leq \frac{j_1-j_2}{2}\\
           \frac13(j_1+2j_2+j_0) \quad j_0 > \frac{j_1-j_2}{2}
          \end{array}\right.
\end{equation}

There are several shortening conditions one can impose. The multiplets $\mathcal{B}_{j_0;j_1,j_2}$ are obtained by imposing
\begin{equation}
 Q^a\ket{\D,j_0,j_1,j_2}^{\text{hw}}=0
\end{equation}
for the three cases  
\begin{align}
 a&=1 & \D&=\frac13(2 j_1+ j_2-j_0) & &\mathcal{B}_{j_0,j_1,j_2}\\
 a&=1,2 & \D&=\frac13(j_2-j_0)  \quad j_1=0 & &\mathcal{B}_{j_0,j_2}\\
 a&=1,2,3 & \D&=-\frac13 j_0 \qquad \quad \, j_1=j_2=0 & &\mathcal{B}_{j_0}
\end{align}
where, compared to \cite{defectABJM} we simplified notation leaving the number of indices to indicate the fraction of supercharges annihilating each multiplet. The conjugate ones are simply given by
\begin{equation}
 \bar Q_a\ket{\D,j_0,j_1,j_2}^{\text{hw}}=0
\end{equation}
for the three cases  
\begin{align}
 a&=3 & \D&=\frac13 ( j_1+2j_2+j_0) & &\bar{\mathcal{B}}_{j_0,j_1,j_2}\\
 a&=2,3 & \D&=\frac13 ( j_1+j_0) \quad j_2=0 & &\bar{\mathcal{B}}_{j_0,j_1}\\
 a&=1,2,3 & \D&=\frac13j_0 \qquad \qquad  j_1=j_2=0 & &\bar{\mathcal{B}}_{j_0}
\end{align}

The remaining multiplets are listed for completeness, but they are not relevant for our setup
\begin{align}
 &\hat{\mathcal{B}}_{j_0,j_1,j_2} & \D&=\frac{j_1+j_2}{2} & j_0&=\frac{j_1-j_2}{2}\\
  &\hat{\mathcal{B}}_{j_0,0,j_2} & \D&=\frac{j_2}{2} & j_0&=\frac{-j_2}{2} & &j_1=0\\
  &\hat{\mathcal{B}}_{j_0,j_1,0} & \D&=\frac{j_1}{2} & j_0&=\frac{j_1}{2} & &j_2=0
\end{align}

We also list the recombination of long multiplets at the unitarity bound. For $j_0< \frac{j_1-j_2}{2}$ the unitarity bound is for $\D=\frac13(2j_1+j_2-j_0)$ and we have
\begin{equation}
 \mathcal{A}^{-\frac13 j_0+\frac23 j_1+\frac13 j_2}_{j_0,j_1,j_2}=\mathcal{B}_{j_0,j_1,j_2} \oplus \mathcal{B}_{j_0+\frac12,j_1+1,j_2}
\end{equation}
Similarly, for $j_0> \frac{j_1-j_2}{2}$ one has
\begin{equation}
 \mathcal{A}^{\frac13 j_0+\frac13 j_1+\frac23 j_2}_{j_0,j_1,j_2}=\bar{\mathcal{B}}_{j_0,j_1,j_2} \oplus \bar{\mathcal{B}}_{j_0-\frac12,j_1,j_2+1}
\end{equation}
For $j_0=\frac{j_1-j_2}{2}$ we have
\begin{equation}
 \mathcal{A}^{j_1+j_2}_{\frac{j_1-j_2}{2},j_1,j_2}=\hat{\mathcal{B}}_{\frac{j_1-j_2}{2},j_1,j_2} \oplus \hat{\mathcal{B}}_{\frac{j_1-j_2}{2}+\frac12,j_1+1,j_2}\oplus \hat{\mathcal{B}}_{\frac{j_1-j_2}{2}-\frac12,j_1+1,j_2+1}\oplus \hat{\mathcal{B}}_{\frac{j_1-j_2}{2},j_1+1,j_2+1}
\end{equation}

For vanishing Dynkin labels the decomposition is different. We first list all short multiplets with vanishing labels as
\begin{align}
 &\{\bar{\mathcal{B}}_{j_0,0,j_2},\mathcal{B}_{j_0,j_2},\hat{\mathcal{B}}_{j_0,0,j_2}\}  & j_1&=0 & j_2&>0 \\
  &\{\mathcal{B}_{j_0,j_1,0},\bar{\mathcal{B}}_{j_0,j_1},\hat{\mathcal{B}}_{j_0,j_1,0}\}  & j_1&>0 & j_2&=0 \\
  &\{\mathcal{B}_{j_0},\bar{\mathcal{B}}_{j_0}\}  & j_1&=0 & j_2&=0 
\end{align}
The decompositions of long multiplet at the unitarity bound for these cases are shown in Table~\ref{longmultdec}.

\begin{table}[htbp]
\begin{center}
\def\arraystretch{1.5}
\begin{tabular}{|c|c|c|}
\hline 
	   	& $j_0<-\frac{j_2}{2}$  & $\mathcal{A}^{\frac13 (j_2-j_0)}_{j_0,0,j_2}=\mathcal{B}_{j_0,j_2}\oplus \mathcal{B}_{j_0+\frac12,1,j_2}$  \\
$j_1=0$		& $j_0>-\frac{j_2}{2}$ & $\mathcal{A}^{\frac13 (2j_2+j_0)}_{j_0,0,j_2}=\bar{\mathcal{B}}_{j_0,0,j_2}\oplus \bar{\mathcal{B}}_{j_0-\frac12,0,j_2+1}$\\
		& $j_0=-\frac{j_2}{2}$ & $\mathcal{A}^{\frac{j_2}{2}}_{-\frac{j_2}{2},0,j_2}=\hat{\mathcal{B}}_{-\frac{j_2}{2},0,j_2}\oplus \hat{\mathcal{B}}_{-\frac{j_2+1}{2},0,j_2+1}\oplus \hat{\mathcal{B}}_{\frac{1-j_2}{2},1,j_2}\oplus \hat{\mathcal{B}}_{-\frac{j_2}{2},1,j_2+1}$\\[1ex]
\hline 
	\To   	& $j_0<\frac{j_1}{2}$  & $\mathcal{A}^{\frac13 (2j_1-j_0)}_{j_0,j_1,0}=\mathcal{B}_{j_0,j_1,0}\oplus \mathcal{B}_{j_0+\frac12,j_1+1,0}$\\
$j_2=0$		& $j_0>\frac{j_1}{2}$ & $\mathcal{A}^{\frac13 (j_1+j_0)}_{j_0,j_1,0}=\bar{\mathcal{B}}_{j_0,j_1}\oplus \bar{\mathcal{B}}_{j_0-\frac12,j_1,1}$\\
		& $j_0=\frac{j_1}{2}$ & $\mathcal{A}^{\frac{j_1}{2}}_{\frac{j_1}{2},j_1,0}=\hat{\mathcal{B}}_{\frac{j_1}{2},j_1,0}\oplus \hat{\mathcal{B}}_{\frac{j_1+1}{2},j_1+1,0}\oplus \hat{\mathcal{B}}_{\frac{j_1-1}{2},j_1,1}\oplus \hat{\mathcal{B}}_{\frac{j_1}{2},j_1+1,1}$\\[1ex]
\hline 
\multirow{ 2}{*}{$j_1=j_2=0$}		&$j_0<0$  & $\mathcal{A}^{-\frac{j_0}{3}}_{j_0,0,0}=\mathcal{B}_{j_0}\oplus \mathcal{B}_{j_0+\frac12,1,0}$\\
			&$j_0>0$  & $\mathcal{A}^{\frac{j_0}{3}}_{j_0,0,0}=\bar{\mathcal{B}}_{j_0}\oplus \bar{\mathcal{B}}_{j_0-\frac12,0,1}$\\[1ex]
\hline
\end{tabular}
\end{center}
\caption{Decomposition of long multiplets into short ones for the case of some vanishing Dynkin labels.}\label{longmultdec}
\end{table}

\section{Supersymmetry transformation of the fields}\label{susyonfields}
\label{app:susytransf}

The supersymmetry transformations of the scalar fields under the preserved supercharges read
\begin{align}
 Q^a Z&=\bar \chi^a_+     &  \bar Q_a Z&=0 &  Q^a \bar Z&=0  &\bar Q_a \bar Z&=i\chi^+_a  \\
 Q^a Y_b&=-\d^a_b \bar \psi_+ &  \bar Q_a Y_b&=i\e_{abc} \bar \chi^c_- & Q^a \bar Y^b&=-\e^{abc}\chi^-_c & \bar Q_a \bar Y^b&=-i\d^b_a\psi^+
\end{align}
and similarly for fermions
\begin{align}
 \bar Q_a \psi^+&=0    & Q^a \psi^+&=-2iD_1 \bar Y^a-\frac{4\pi i}{k}[ \bar Y^a l_B-\hat{l}_B \bar Y^a]  \\
 Q^a \psi^-&=-2D\bar Y^a  &\bar Q_a \psi^-&=-\frac{8\pi }{k}\e_{abc}\bar Y^b Z \bar Y^c  \\
   \bar Q_a \chi^+_b&=2i\e_{abc}\bar D \bar Y^c & Q^a \chi^+_b&=2i\d^a_b D_1 \bar Z+\frac{8\pi i}{k} [\bar Z\Lambda^a_b-\hat\Lambda^a_b\bar Z]     \\
  Q^a \chi^-_b&=2\d^a_b D \bar Z &\bar Q_a \chi^-_b&=-2\e_{abc}D_1 \bar Y^c-\frac{4\pi }{k}\e_{acd}[\bar Y^c  \Theta^d_b-\hat\Theta^d_b \bar Y^c]  \\
   Q^a \bar \psi_+&=0    & \bar Q_a \bar \psi_+&=2 D_1  Y_a+\frac{4\pi }{k}[ Y_a \hat{l}_B-l_B  Y_a]  \\
 \bar Q_a \bar \psi_-&=2 i\bar D Y_a  & Q^a \bar \psi_-&=-\frac{8\pi i}{k}\e^{abc} Y_b \bar Z  Y_c  \\
    Q^a \bar \chi_+^b&=2\e^{abc} D  Y_c & \bar Q_a \bar \chi_+^b&=-2\d^b_a D_1 Z-\frac{8\pi }{k} [Z \hat\Lambda_a^b-\Lambda_a^b Z]     \\
  \bar Q_a \bar \chi_-^b&=-2i\d^b_a \bar D Z & Q^a \bar \chi_-^b&=-2i\e^{abc}D_1 Y_c-\frac{4\pi i}{k}\e^{acd}[  Y_c \hat\Theta_d^b-\Theta_d^b Y_c]
\end{align}
where we used the definitions
\begin{align}
  D&=D_2-i D_3 & \bar D&=D_2+i D_3
  \end{align}
and the entries of the supermatrices
\begin{align}
\begin{pmatrix} \L_a^b  & 0\\
             0 & \hat \L_a^b\end{pmatrix}&=\begin{pmatrix}Y_a\bar Y^b+\frac12\d^b_a l_B   & 0\\
             0 & \bar Y^b Y_a+\frac12\d^b_a \hat{l}_B \end{pmatrix} \\
\begin{pmatrix} \Theta_a^b  & 0\\
             0 & \hat \Theta_a^b\end{pmatrix}&=\begin{pmatrix} Y_a\bar Y^b-\d^b_a (Y_c\bar Y^c+Z \bar Z)  & 0\\
             0 & \bar Y^b Y_a-\d^b_a (\bar Y^c Y_c+\bar Z Z)\end{pmatrix}\\
 \begin{pmatrix} l_B  & 0\\
             0 & \hat{l}_B\end{pmatrix}&=\begin{pmatrix} (Z \bar Z-Y_a \bar Y^a)  & 0\\
             0 &  (\bar Z Z -\bar Y^a Y_a )\end{pmatrix}
\end{align}
Notice that, due to the last identity the bosonic part of the superconnection reads
\begin{align}
 \mathcal{L}_B=\frac{2\pi i}{k} \begin{pmatrix} l_B  & 0\\
             0 & \hat{l}_B\end{pmatrix}
\end{align}
Finally we can list the transformation properties of the gauge fields
\begin{align}
 Q^a A_1&=\frac{2\pi i}{k} ( \bar \psi_+ \bar Y^a-\bar \chi^a_+ \bar Z- \e^{abc} Y_b \chi_c^-) & \bar Q_a A_1&=\frac{2\pi }{k} (Z\chi_a^+-Y_a\psi^+ + \e_{abc} \bar \chi^b_- \bar Y^c) \\
  Q^a A&=0 &  \bar Q_a A&=-\frac{4\pi i }{k} ( Y_a \psi^- - Z \chi_a^- + \e_{abc} \bar \chi^b_+ \bar Y^c)\\
  Q^a \bar A&=\frac{4\pi }{k} ( \bar \psi_- \bar Y^a-\bar \chi^a_- \bar Z+\e^{abc} Y_b \chi_c^+) &  \bar Q_a \bar A&=0\\
  Q^a \hat A_1&=\frac{2\pi i}{k} ( \bar Y^a \bar \psi_+ - \bar Z \bar \chi^a_+ - \e^{abc} \chi_c^- Y_b ) & \bar Q_a \hat A_1&=\frac{2\pi }{k} (\chi_a^+ Z-\psi^+  Y_a + \e_{abc}  \bar Y^c \bar \chi^b_-) \\
  Q^a \hat A&=0 &  \bar Q_a \hat A&=-\frac{4\pi i }{k} (  \psi^- Y_a- \chi_a^- Z+\e_{abc}  \bar Y^c \bar \chi^b_+)\\
  Q^a \bar {\hat A}&=\frac{4\pi }{k} (  \bar Y^a \bar \psi_- - \bar Z \bar \chi^a_-  +  \e^{abc}  \chi_c^+ Y_b) &  \bar Q_a \bar{\hat A}&=0
\end{align}

To check the closure of these transformations and to use them on local operators it is important to keep in mind the equations of motion.
For the gauge field we are interested in the components  $\mathcal{F}= \mathcal{F}_{21} -i  \mathcal{F}_{31}$ and   $\bar{\mathcal{F}}=  \mathcal{F}_{21} +i  \mathcal{F}_{31}$ of the field strength. In particular we focus on the first one, which respects the equation
\begin{equation}\label{Feqmot}
 \mathcal{F} =\frac{2\pi i}{k} \begin{pmatrix} Z\overleftrightarrow{D}\bar Z+ Y_a \overleftrightarrow{D}\bar Y^a+\bar\psi_+\psi^- +\bar\chi_+^a\chi^-_a & 0\\
            0& -\bar Z\overleftrightarrow{D}Z- \bar Y^a \overleftrightarrow{D}Y_a -\psi^-\bar\psi_+ -\chi^-_a\bar\chi_+^a
            \end{pmatrix} 
\end{equation}
where the operator $\overleftrightarrow{D}$ has the usual definition $Z\overleftrightarrow{D}\bar Z\equiv ZD\bar Z-DZ\bar Z$.
For the fermions we need the equation
\begin{equation}
 \slashed{D}\psi_J=\frac{2\pi}{k} \left(\bar C^I C_I\psi_J-\psi_J C_I \bar C^I +2 \psi_I C_J \bar C^I-2 \bar C^I C_J \psi_I + 2\e_{ILKJ} \bar C^I \bar \psi^L \bar C^K\right)
 \end{equation}
whose projection yields (we list just the components we needed for our computations)
\begin{align}\label{fermeq}
 D\psi^+ = i D_1 \psi^- &+ \frac{2\pi i}{k}\left( \hat l_B \psi^- - \psi^- l_B +2\bar Y^a Z\chi^-_a -2 \chi^-_a Z \bar Y^a-2\bar Y^a \bar \chi_+^b \bar Y^c \e_{abc}\right)\\
 D \chi _a^+=i D_1\chi ^-_a &+\frac{2 \pi i}{k}  \left(\chi ^-_b \O_a^b- \hat \O_a^b \chi ^-_b-2 \bar{Z} Y_a \psi ^-+2 \psi ^-
   Y_a \bar{Z}+\right)\\
   &+\frac{4 \pi i}{k}\epsilon _{a c d} \left(\bar{Y}^c \bar{\psi }_+ \bar{Y}^d+\bar{Y}^d \bar{\chi }_+^c \bar{Z}-\bar{Z} \bar{\chi }_+^c \bar{Y}^d\right)
   \end{align}
with 
\begin{equation}
 \O_a^b= \Theta_a^b+ \Lambda_a^b-\frac12 \d_a^b l_B.
\end{equation}

\section{Details on the analytic bootstrap}\label{detailsbootstrap}
As we highlighted in the main text, the relation between \eqref{4-point-1-primary} and \eqref{4-point-2-primary} is not simply a consequence of crossing as in higher dimensional CFTs. Here, we show that, in perturbation theory one can establish a relation between these two correlators, finding the expression \eqref{ansatzh}. In free theory, we pointed out in section \ref{leadingorder} that the $s$-channel exchanged operators $[\mathbb{F}\bar{\mathbb{F}}]_n$ have charge $n$ under the parity symmetry described at the end of section \ref{superblocks} leading to the simple relation $f^{(0)}(\chi)=h^{(0)}(\chi)$. We want to consider a small perturbation of this solution such that the exchanged operators still have charge $n$, but their dimension receives an anomalous contribution $\Delta_n=1+n+\e \gamma^{(1)}_n$. The two $s$-channel expansions \eqref{superblock} and \eqref{hexpansion} at first order read
\begin{align}
 f^{(1)}(z)&=\sum_{n} (-z)^{n+1} \left(c^{(0)}_n \gamma^{(1)}_n \partial_\Delta F_\Delta(z)|_{\Delta=1+n}+ F_{n+1}(z) (c_n^{(0)} \gamma^{(1)}_n \log (-z)+c^{(1)}_n)\right) \label{oneloopexpansionf} \\
 h^{(1)}(\chi)&=\sum_{n} \chi^{n+1} \left(\tilde c^{(0)}_n \gamma^{(1)}_n \partial_\Delta F_\Delta(\chi)|_{\Delta=1+n}+ F_{n+1}(\chi) (\tilde c_n^{(0)} \gamma^{(1)}_n \log (\chi)+\tilde c^{(1)}_n)\right) \label{oneloopexpansionh}
\end{align}
where $F_{\Delta}(z)={}_2 F_{1} (\Delta,\Delta,2\Delta-3,z)$. We can use the relation
\begin{align}
 c^{(l)}_n=(-1)^{1+n} \tilde c^{(l)}_n
\end{align}
which only depends on the quantum number of the exchanged operator under parity and therefore it remains true perturbatively. Therefore, we immediately see that the two expansions \eqref{oneloopexpansionf} and \eqref{oneloopexpansionh} are mapped to each other by the transformation $\chi\to z$, up to the sign of the argument of the logarithm. Starting from the ansatz \eqref{ansatz} for $\hat{f}(\chi)$ one only needs to use \eqref{fhatdef} and \eqref{z-chi-rel} to obtain
\begin{align}
 f^{(1)}(z)=\frac{z}{z-1}\left[r\left(\tfrac{1}{1-z}\right)\log (-z) - \left[r\left(\tfrac{z}{z-1}\right)+r\left(\tfrac{1}{1-z}\right)\right]\log (1-z)+q\left(\tfrac{z}{1-z}\right)\right]
\end{align}
From this expression, using the argument above we immediately find
\begin{align}
 h^{(1)}(\chi)=\frac{\chi}{\chi-1}\left[r\left(\tfrac{1}{1-\chi}\right)\log (\chi) - \left[r\left(\tfrac{\chi}{\chi-1}\right)+r\left(\tfrac{1}{1-\chi}\right)\right]\log (1-\chi)+q\left(\tfrac{\chi}{1-\chi}\right)\right]
\end{align}
from which \eqref{ansatzh} descends immediately.

We now show that the bootstrap problem for the first order perturbation at strong coupling has an infinite number of solutions parametrized by the coefficients $q_l$ in \eqref{qexpansion}.  We start by considering $\hat{f}^{(1)}(\chi)$ in \eqref{ansatz} in the limit $\chi\to 0$. As we mentioned in the main text, we need to have a cancellation between the poles in $r(\chi)\log(1-\chi)$ and those in $ q(\chi)$ to have a regular expansion for $\hat{f}^{(1)}(\chi)$. We can then expand
\begin{align}
 r(\chi)\log(1-\chi)&=-\sum_{m=-M_1}^{M_2}\sum_{p=1}^{\infty} \frac{r_m}{p} \chi^{m+p}\\
 q(\chi)&=\sum_{l=-L_1}^{L_2} \sum_{q=0}^{\infty} (-1)^q q_l \begin{pmatrix} l\\q\end{pmatrix}\chi^{q+l}
\end{align}
After shifting the argument of the sums we find
\begin{align}
 r(\chi)\log(1-\chi)&=-\sum_{p=-M_1+1}^{\infty}\sum_{m=-M_1}^{\min(p-1, M_2)} \frac{r_m}{p-m} \chi^{p}\\
 q(\chi)&=\sum_{q=-L_1}^{\infty}\sum_{l=-L_1}^{\min(q,L_2)}  (-1)^{q-l} q_l \begin{pmatrix} l\\q-l\end{pmatrix}\chi^{q}
\end{align}
The cancellation of the singular behaviour in the sum of these two functions requires that $M_1=L_1+1$ and that
\begin{align}
 \sum_{m=-L_1-1}^{p-1}  \frac{r_m}{p-m} =\sum_{l=-L_1}^{p}  (-1)^{p-l} q_l \begin{pmatrix} l\\p-l\end{pmatrix}
\end{align}
valid for $-L_1\leq p\leq -1$. This is a matrix equation for the coefficients $r_m$ and the system has maximal rank because the matrix multiplying $r_m$ is upper triangular. Therefore this allows to fix all the coefficients $r_m$ for $-M_1\leq m \leq -2$ in terms of the coefficients $q_l$.

Similarly, we can start from the expression of $\hat h(\chi)$ given in \eqref{ansatzh} and consider the expansion around $\chi=1$. We have
\begin{align}
 r\left(\tfrac{1}{1-\chi}\right) \log \chi=-\sum_{m=-M_1}^{M_2} \sum_{p=-m+1}^{\infty} \frac{r_m}{p+m} (1-\chi)^{p}\\
 q\left(\tfrac{\chi}{\chi-1}\right) =\sum_{l=-L_1}^{L_2} \sum_{q=-2l}^{\infty} q_l (-1)^{l+q} \begin{pmatrix} l \\ q+2l \end{pmatrix} (1-\chi)^q 
\end{align}
By switching the order of the sums we get
\begin{align}
 r\left(\tfrac{1}{1-\chi}\right) \log \chi=-\sum_{p=-M_2+1}^{\infty} \sum_{m=\max(-M_1,-p+1)}^{M_2}  \frac{r_m}{p+m} (1-\chi)^{p}\\
 q\left(\tfrac{\chi}{\chi-1}\right) = \sum_{q=-2L_2}^{\infty} \sum_{l=\max(-L_1,[-\frac{q}{2}])}^{L_2} q_l (-1)^{l+q} \begin{pmatrix} l \\ q+2l \end{pmatrix} (1-\chi)^q 
\end{align}
In order for the poles to cancel in the sum of these two terms we need to have $M_2=2L_2+1$ and 
\begin{align}
 \sum_{m=-p+1}^{2L_2+1}  \frac{r_m}{p+m}=\sum_{l=[-\frac{p}{2}]}^{L_2} q_l (-1)^{l+p} \begin{pmatrix} l \\ p+2l \end{pmatrix}
\end{align}
for $-2L_2\leq p \leq 0$. This is another system of maximal rank, whose solution allows to determine all the coefficients $r_m$ for $0\leq m\leq M_2$ in terms of the coefficients $q_l$. Therefore we are left with two unfixed coeffcients $r_0$ and $r_{-1}$. To fix these two coefficients one can look at the term proportional to $\log(1-\chi)$ in \eqref{ansatzh}. In the $\chi\to 1$ limit that term identifies the anomalous dimensions of the operators exchanged in the chiral-chiral channel. We can then expand its coefficient around $\chi\to 1$
\begin{align}
 r\left(\tfrac{\chi}{\chi-1}\right)+ r\left(\tfrac{1}{1-\chi}\right)=\sum_{m=-M_1}^{M_2} \frac{r_{m}}{(1-\chi)^m}+ \sum_{m=-M_1}^{M_2} \sum_{n=0}^{\infty} \frac{r_m}{(\chi-1)^{m-n}} \begin{pmatrix} m\\n\end{pmatrix}
\end{align}
Requiring that this expansion starts at $(1-\chi)^3$ in the $\chi\to 1$ limit leads to several relations among the coefficients $r_m$. Nevertheless, only two such relations are independent and they allow to fix $r_0$ and $r_{-1}$ in terms of the other coefficients (and therefore in terms of $q_l$). These relations are
\begin{align}
 2r_{0}+\sum_{m\neq0} r_m&=0 & r_{-1}=2r_{-2}
\end{align}
To sum up, we have found that all the coefficients $r_m$ in the expansion \eqref{rexpansion} are fixed in terms of the $q_l$ in \eqref{qexpansion}, leaving us with infinitely many solutions parametrized by these coefficients.

 \section{D-functions}
 \label{app:D-functions}
 
Tree-level correlators obtained via contact diagrams may be written in terms of $D$-functions \cite{Liu:1998ty,DHoker:1999kzh, Dolan:2000ut}, defined in the general  case of AdS$_{d+1}$ as 
\be \label{D-function}
\!\!\!\!\!\!\!\!
D_{\Delta_1\Delta_2\Delta_3\Delta_4}(x_1,x_2,x_3,x_4) =\! \!\int \!\!\frac{dz d^dx}{z^{d+1}} 
\tilde{K}_{\Delta_1}\!(z,x;x_1) \tilde{K}_{\Delta_2}\!(z,x;x_2) \tilde{K}_{\Delta_3}\!(z,x;x_3) \tilde{K}_{\Delta_4}\!(z,x;x_4)
 \ee
 in term of the bulk-to-boundary propagator in $d$ dimensions
\be
K_{\Delta}(z,x;x') = {\cal C}_{\Delta} \Big[\frac{z}{z^2+(x-x')^2}\Big]^\Delta \equiv  {\cal C}_{\Delta}\,  \tilde{K}_{\Delta}(z,x;x')\,,
\ee
where ${\cal C}_{\Delta}$ is defined in~\eqref{bulk-to-boundary}. 
When dealing with derivatives in the vertices, the following identity is useful 
\bal
&g^{\m\n}\partial_\m \tilde{K}_{\Delta_1}(z,x;x_1)\ \partial_\n\tilde{K}_{\Delta_2}(z,x;x_2) 
\\   & \qquad = 
\Delta_1\Delta_2
\left[\tilde{K}_{\Delta_1}(z,x;x_1)\tilde{K}_{\Delta_2}(z,x;x_2)-2x_{12}^2 \tilde{K}_{\Delta_1+1}(z,x;x_1)\tilde{K}_{\Delta_2+1}(z,x;x_2)\right]\ \,,
 \label{identityder}
\eal
with  $g^{\m\n}= {z^2}\delta^{\m\n}$ and $\del_\m=(\del_z,\del_r)$, \ $r=0,1, 2, ...,d-1$. 
Reduced $D$-functions are defined via \eqref{D-function} as \cite{Dolan:2003hv}  ($\Sigma \equiv \frac{1}{2}\sum_i \Delta_i$)
\be \label{Dbar}
\!\!\!\!
D_{\Delta_1\Delta_2\Delta_3\Delta_4}= 
\frac{\pi^{d\ov 2}\Gamma\left(\Sigma-{d\ov2}\right)}{2\, \Gamma\left(\Delta_1\right)\Gamma\left(\Delta_2\right)\Gamma\left(\Delta_3\right)\Gamma\left(\Delta_4\right)}
\frac{x_{14}^{2(\Sigma-\Delta_1-\Delta_4)} x_{34}^{2(\Sigma-\Delta_3-\Delta_4)}}
{x_{13}^{2(\Sigma-\Delta_4)} x_{24}^{2\Delta_2}}\bar{D}_{\Delta_1\Delta_2\Delta_3\Delta_4}(u,v)
\ee
and depend only on the cross-ratios $u=\frac{x_{12}x_{34}}{x_{13}x_{24}}\,,  v=\frac{x_{14}x_{23}}{x_{13}x_{24}}$.  Their explicit expression in terms a Feynman 
parameter integral reads
\begin{equation}
\bar{D}_{\Delta_1\Delta_2\Delta_3\Delta_4}(u,v)=
\int d\alpha d\beta d\gamma\  \delta(\alpha+\beta+\gamma-1)\ 
\alpha ^{\Delta _1-1} \beta ^{\Delta _2-1} \gamma ^{\Delta _3-1} 
\frac{\Gamma \left(\Sigma-\Delta _4\right) \Gamma\left(\Delta_4\right)}
{\big(\alpha  \gamma + \alpha  \beta\, u  + \beta  \gamma\, v\big)^{\Sigma-\Delta _4}}\,,
\label{Dbar-integral}
\end{equation}
while in  $d=1$ as usual they only depend on the single variable   $\chi$ ($u=\chi^2$, $v=(1-\chi)^2$). 
Explicit expressions for the $\bar{D}$-functions appearing in this paper read
\begin{eqnarray}\label{Dbar-explicit}
\bar D_{1,1,1,1}&=&-\frac{2 \log (1-\chi )}{\chi }-\frac{2 \log (\chi )}{1-\chi }\\
\bar{D}_{2,2,1,1} &=&-\frac{(\chi +2) \log (1-\chi )}{3 \chi ^3}+\frac{1}{3 (1-\chi ) \chi ^2}+\frac{\log (\chi )}{3 (1-\chi )^2}\\
\bar D_{1,2,2,1}&=&   \frac{\log (1-\chi )}{3 \chi ^2}+\frac{1}{3 (1-\chi )^2 \chi }-\frac{(\chi -3) \log (\chi
   )}{3 (\chi -1)^3}\\
\bar{D}_{1,2,1,2} &=& -\frac{(2 \chi +1) \log (1-\chi )}{3 \chi ^2}-\frac{1}{3 (1-\chi ) \chi }+\frac{(2 \chi -3) \log (\chi )}{3 (1-\chi )^2}\\
\bar{D}_{2,2,2,2} &=&-\frac{2 \left(\chi ^2-\chi +1\right)}{15 (1-\chi )^2 \chi ^2}+\frac{\left(2 \chi ^2-5 \chi +5\right) \log (\chi )}{15 (\chi -1)^3}-\frac{\left(2 \chi ^2+\chi +2\right) \log (1-\chi )}{15 \chi ^3}\\
\bar{D}_{2,3,1,2}&=&-\frac{(\chi  (3 \chi +4)+3) \log (1-\chi )}{15 \chi ^4}-\frac{\chi  (3 \chi -8)+3}{15 (\chi -1)^2 \chi ^3}+\frac{(3 \chi -5) \log (\chi )}{15 (\chi -1)^3}\\
\bar{D}_{2,3,2,1} &=&\frac{(2 \chi +3) \log (1-\chi )}{15 \chi ^4}+\frac{2 (\chi -1) \chi -3}{15 (\chi -1)^3 \chi ^3}+\frac{(5-2 \chi ) \log (\chi )}{15 (\chi -1)^4}\\
\bar{D}_{3,3,1,1} &=& -\frac{2 (\chi  (\chi +3)+6) \log (1-\chi )}{15 \chi ^5}+\frac{3-\chi  (2 \chi +3)}{15 (\chi -1)^2 \chi ^4}+\frac{2 \log (\chi )}{15 (\chi -1)^3}\\
\bar{D}_{2,3,2,3}&=&\frac{\left(12 \chi ^3-42 \chi ^2+56 \chi -35\right) \log (\chi )}{105 (\chi -1)^4}+\frac{-24 \chi ^4+48 \chi ^3+5 \chi ^2-29 \chi +18}{210 (\chi -1)^3 \chi ^3}\nonumber\\
&&+\frac{\left(-12 \chi ^3-6 \chi ^2-8 \chi -9\right) \log (1-\chi )}{105 \chi ^4}\\\nonumber
\bar{D}_{2,3,3,2}&=&\frac{\left(9 \chi ^2+10 \chi +9\right) \log (1-\chi )}{105 \chi ^4}+\frac{\left(-9 \chi ^3+35 \chi ^2-49 \chi +35\right) \log (\chi )}{105 (\chi -1)^5}\nonumber\\
&&+\frac{18 \chi ^4-43 \chi ^3+26 \chi ^2-43 \chi +18}{210 (\chi -1)^4 \chi ^3}\\
\bar{D}_{3,3,2,2}&=&\frac{\left(9 \chi ^2-28 \chi +28\right) \log (\chi )}{105 (\chi -1)^4}+\frac{\left(-9 \chi ^3-8 \chi ^2-6 \chi -12\right) \log (1-\chi )}{105 \chi ^5}\nonumber\\
&&+\frac{-18 \chi ^4+29 \chi ^3-5 \chi ^2-48 \chi +24}{210 (\chi -1)^3 \chi ^4}\\
\bar{D}_{3,3,3,3}&=&\frac{\left(8 \chi ^4-36 \chi ^3+64 \chi ^2-56 \chi +28\right) \log (\chi )}{105 (\chi -1)^5}+\frac{\left(-8 \chi ^4-4 \chi ^3-4 \chi ^2-4 \chi -8\right) \log (1-\chi )}{105 \chi ^5}\nonumber\\
&&+\frac{-24 \chi ^6+72 \chi ^5-74 \chi ^4+28 \chi ^3-74 \chi ^2+72 \chi -24}{315 (\chi -1)^4 \chi ^4}\,.
\label{Dbar-explicit-end}
\end{eqnarray}
Further expressions are found through the identities in~\cite{Dolan:2003hv} 
\begin{eqnarray}
\bar D_{1,1,2,2}&=&\chi^2 \bar{D}_{2,2,1,1} = \frac{\chi ^2 \log (\chi )}{3 (1-\chi )^2}+\frac{1}{3 (1-\chi )}-\frac{(\chi +2) \log
	(1-\chi )}{3 \chi }\\
\bar{D}_{2,1,2,1}&=&\bar{D}_{1,2,1,2}\\
\bar{D}_{2,1,1,2} &=&(1-\chi)^2\bar{D}_{1,2,2,1}=\frac{(1-\chi )^2 \log (1-\chi )}{3 \chi ^2}+\frac{1}{3 \chi }+\frac{(\chi -3) \log (\chi )}{3 (1-\chi )}\\
\bar{D}_{3,2,2,1}&=&=\bar{D}_{2,3,1,2}=-\frac{(\chi  (3 \chi +4)+3) \log (1-\chi )}{15 \chi ^4}-\frac{\chi  (3 \chi -8)+3}{15 (\chi -1)^2 \chi ^3}+\frac{(3 \chi -5) \log (\chi )}{15 (\chi -1)^3}\\
\bar{D}_{3,2,1,2} &=&(1-\chi)^2\bar{D}_{2,3,1,2}=\frac{(2 \chi +3) (\chi -1)^2 \log (1-\chi )}{15 \chi ^4}+\frac{2 (\chi -1) \chi -3}{15 (\chi -1) \chi ^3}+\frac{(5-2 \chi ) \log (\chi )}{15 (\chi -1)^2}\\
\bar{D}_{2,2,3,3}&=&\chi^2\bar{D}_{3,3,2,2}=\frac{\left(-9 \chi ^3-8 \chi ^2-6 \chi -12\right) \log (1-\chi )}{105 \chi ^3}+\frac{-18 \chi ^4+29 \chi ^3-5 \chi ^2-48 \chi +24}{210 (\chi -1)^3 \chi ^2}\nonumber\\
&&\qquad\qquad\qquad+\frac{\left(9 \chi ^4-28 \chi ^3+28 \chi ^2\right) \log (\chi )}{105 (\chi -1)^4}\\
\bar{D}_{3,2,2,3} &=&(1-\chi)^2\bar{D}_{2,3,3,2}=\frac{\left(-9 \chi ^3+35 \chi ^2-49 \chi +35\right) \log (\chi )}{105 (\chi -1)^3}+\frac{18 \chi ^4-43 \chi ^3+26 \chi ^2-43 \chi +18}{210 (\chi -1)^2 \chi ^3}\nonumber\\
&&\qquad\qquad\qquad+\frac{\left(9 \chi ^4-8 \chi ^3-2 \chi ^2-8 \chi +9\right) \log (1-\chi )}{105 \chi ^4}
\end{eqnarray}



\bibliographystyle{nb}
\bibliography{Ref_Witten}

\end{document}